\documentclass[10pt,journal,compsoc]{IEEEtran}
\usepackage{tabularx}
\usepackage{times}
\usepackage{epsfig}
\usepackage{booktabs}
\usepackage{changepage}
\usepackage{graphicx}
\usepackage{subcaption}
\usepackage{amsmath}
\usepackage[normalem]{ulem}
\usepackage{caption}
\usepackage{cite}
\usepackage{xcolor}
\usepackage{makecell}
\usepackage{pgf-pie} 
\usepackage{adjustbox} 
\usepackage{chngcntr}
\usepackage{enumitem}
\usepackage{hyperref}
\usepackage{array}
\usepackage{booktabs,arydshln} 
\usepackage{amssymb,enumitem}
\usepackage{tikz}
\usetikzlibrary{trees,tikzmark,calc,positioning,shapes.geometric}
\usetikzlibrary{matrix, arrows.meta}
\usepackage{multirow}
\usepackage{colortbl}
\usepackage{orcidlink}
\definecolor{lightblue}{rgb}{0.88,0.91,1.0}

\setlist[itemize,1]{label=\tiny$\blacksquare$,left=0em,labelsep=1em}
\newcolumntype{C}[1]{>{\centering\arraybackslash}p{#1}}
\makeatletter
\def\thesubsection{\thesection.\@arabic\c@subsection}
\makeatother

\begin{document}
\title{Computational Analysis of Stress, Depression and Engagement in Mental Health: A Survey}

\author{
    Puneet~Kumar
        \orcidlink{0000-0002-4318-1353},
        \IEEEmembership{Member,~IEEE},
    Alexander~Vedernikov
        \orcidlink{0000-0003-2127-0934},
        \IEEEmembership{Member,~IEEE},
    Yuwei~Chen
        \orcidlink{0000-0003-0148-3609},
    Wenming~Zheng
        \orcidlink{},
        \IEEEmembership{Senior~Member,~IEEE},
    Xiaobai Li*
        \orcidlink{0000-0002-4318-1353},        \IEEEmembership{Senior~Member,~IEEE}~\thanks{*Corresponding Author.}
    \thanks{P. Kumar and A. Vedernikov are with the Center for Machine Vision and Signal Analysis, University of Oulu, Finland. Email: puneet.kumar@oulu.fi, aleksandr.vedernikov@oulu.fi.}
    \thanks{Y. Chen is with Hangzhou Institute for Advanced Study, University of Chinese Academy of Sciences, Hangzhou, China. Email: yuwei.chen@ucas.ac.cn.}
    \thanks{W. Zheng is with the Key Laboratory of Child Development and Learning Science (Southeast University), Ministry of Education and also with the School of Biological Science and Medical Engineering, Southeast University, Nanjing 210096, China. Email: wenming\_zheng@seu.edu.cn}
    \thanks{X. Li is with the State Key Laboratory of Blockchain and Data Security, Zhejiang University, Hangzhou, China and the Center for Machine Vision and Signal Analysis, University of Oulu, Finland. E-mail: xiaobai.li@zju.edu.cn}
    \thanks{Manuscript Received: Mar 2025.} 
}

\markboth{Journal of \LaTeX\ Class Files,~Vol.~14, No.~8, August~2015}%
{Shell \MakeLowercase{\textit{P. Kumar et al.}}: Bare Demo of IEEEtran.cls for IEEE Journals}

\IEEEtitleabstractindextext{
\begin{abstract}
Analysis of stress, depression and engagement is less common and more complex than that of frequently discussed emotions such as happiness, sadness, fear and anger. The importance of these psychological states has been increasingly recognized due to their implications for mental health and well-being. Stress and depression are interrelated and together they impact engagement in daily tasks, highlighting the need to explore their interplay. This survey is the first to simultaneously explore computational methods for analyzing stress, depression and engagement. We present a taxonomy and timeline of the computational approaches used to analyze them and we discuss the most commonly used datasets and input modalities, along with the categories and generic pipeline of these approaches. Subsequently, we describe state-of-the-art computational approaches, including a performance summary on the most commonly used datasets. Following this, we explore the applications of stress, depression and engagement analysis, along with the associated challenges, limitations and future research directions.
\end{abstract}

}

\maketitle
\IEEEpeerreviewmaketitle
\section{Introduction}\label{sec:intro}
\IEEEPARstart{A}{ffective} Computing involves the development of computational approaches to analyze a broad spectrum of psychological states \cite{kim2018building}. \textit{Psychological State} is a broad term encompassing various mental conditions related to affect and cognition \cite{kachele2014fusion}. \textit{Affect} refers to the experience of feeling or emotion, including broader concepts such as \textit{Emotion}, \textit{Mood}, \textit{Sentiment} and \textit{Opinion}. These aspects collectively characterize how individuals experience and express their emotional states. In contrast, \textit{Cognition} involves the mental processes of acquiring knowledge and understanding through thought, experience and the senses \cite{arnsten2015stress}. This includes functions like perception, memory and judgment, crucial for processing and interpreting information. \textit{Emotion} is a manifestation of affect marked by complex mental states and physiological responses, with theories that categorize emotions through dimensions like valence and arousal or into discrete classes such as happiness, sadness, fear and anger \cite{ekman1992argument, ps2017emotion}. \textit{Mood} is a more lasting but less intense expression of affect, whereas emotions give rise to \textit{Sentiments} over time, which are basic mental attitudes \cite{akhtar2019all}. The sentiments subsequently lead to \textit{Opinions}, which are personal interpretations shaped by one's experience \cite{munezero2014they}. \vspace{.025in}  

\begin{figure}[!t]
  \centering
  \includegraphics[width=.3\textwidth]{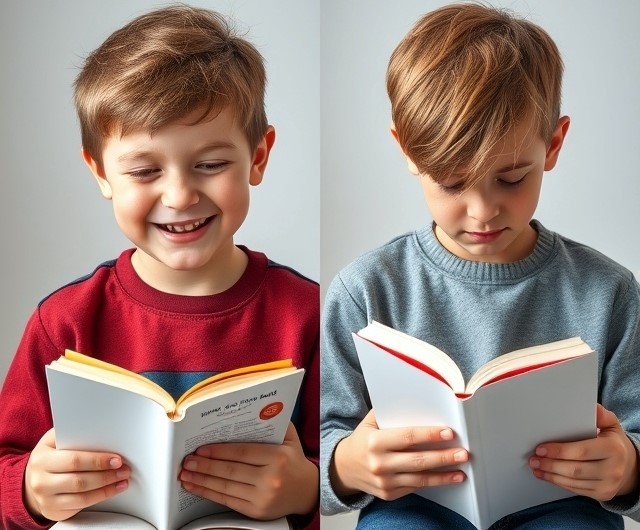}\vspace{-.05in}
  \caption{{Illustration showing the interaction between basic emotions and psychological states: the left image shows a boy happy and engaged with his reading, while the right image shows him sad but still engaged, demonstrating that psychological states can coexist with different emotions. This image was created with \href{https://openai.com/dall-e-2}{DALL·E 2}.\vspace{-.15in}}} 
  \label{fig:intro}
\end{figure}

There are six basic emotions: anger, surprise, disgust, happiness, fear and sadness, as proposed by Paul Ekman \cite{ekman1992argument}. These emotions are extensively studied and universally recognized in different cultures and ethnicities. In contrast, there are complex psychological states such as stress, guilt, depression, shame, pride, curiosity, empathy, envy, engagement, etc., that are not as frequently explored in the literature \cite{draghi2001rethinking}. {Fig. \ref{fig:intro} demonstrates how basic emotions interact with complex psychological states, illustrating that various psychological states can coexist with diverse emotional contexts.} This survey explores a broad range of psychological states that extend beyond basic emotions, with a focus on those relevant to mental health analysis. Appraisal theories, as proposed by Scherer \cite{scherer2009dynamic} and Roseman \cite{roseman1990appraisals}, offer a richer framework for understanding the complexities of emotional states beyond simple discrete categories. According to these models, emotions are appraised through multiple dimensions, which influence how they are perceived and experienced. Furthermore, embodying emotion theory \cite{niedenthal2007embodying} suggests that emotions involve bodily responses, while constructivist theory \cite{barrett2017emotions} argues that emotions are constructed from core psychological systems, rather than triggered by external events. This perspective advocates a deeper exploration of less explored psychological states. \vspace{.025in}   

This survey reviews computational approaches for analyzing stress, depression and engagement, which are interrelated and play a significant role in mental health analysis \cite{wrosch2002health}. {Stress disrupts attentional networks and motivation \cite{arnsten2015stress}, while the lack of interest associated with depression further compromises an individual’s ability to remain engaged \cite{pizzagalli2014depression}. Stress and depression reduce daily efficiency and task performance \cite{gurel2020automatic, maydych2019interplay}.} Several computational studies have highlighted the correlation among stress, depression and engagement. For instance, Pizzagalli et al. \cite{pizzagalli2014depression} have presented an integrated model in which stress disrupts reward processing and leads to anhedonia which is a core mechanism linking stress to depression. In another work, Slavich and Irwin \cite{slavich2014stress} have proposed a social signal transduction theory that explains how stress-induced inflammation contributes to the development of major depressive disorder, thereby impairing cognitive functions essential for maintaining engagement. Moreover, a longitudinal study by Innstrand et al. \cite{innstrand2012longitudinal} demonstrated that reduced work engagement is significantly associated with increased symptoms of depression and anxiety, underscoring the dynamic interplay among these psychological states.\vspace{.025in}

Neurophysiological studies have also indicated the interconnected nature of stress, depression and engagement in cognitive and emotional processes, underscoring their importance for mental health and well-being \cite{lupien2009effects}. Chronic stress has been shown to induce key neurobiological changes that affect synaptic integrity and neurotransmission in crucial regions such as the limbic system and prefrontal networks, ultimately leading to neuroendocrine dysfunction and an increased vulnerability to depression, as detailed by Krishnan et al. \cite{krishnan2008molecular} and expanded by Lupien et al. \cite{lupien2009effects}. Similarly, Arnsten \cite{arnsten2015stress} and Liston et al. \cite{liston2009psychosocial} have highlighted that stress negatively impacts the prefrontal cortex, impairing cognitive functions such as executive control and attention which are essential for sustaining engagement in demanding tasks. In addition, Pizzagalli et al. \cite{pizzagalli2014depression} have shown that stress-related changes in the brain’s reward circuits, particularly in the ventral striatum and medial prefrontal cortex, result in anhedonia and motivational deficits typical of depression, further reducing engagement by diminishing responsiveness to rewarding stimuli. Heller \cite{heller2009reduced} has noted that depression is correlated with a decreased ability to maintain activation in frontostriatal networks, which are crucial for generating and sustaining positive emotions and engagement. Lastly, Slavich and Irwin \cite{slavich2014stress} have proposed a model linking psychosocial stress to neuroimmune alterations affecting mood- and cognition-related neural circuits, thereby further bridging stress and depression on a neurophysiological level.\vspace{.025in} 

While previous surveys have individually addressed stress (\cite{giannakakis2019review, nvemcova2020multimodal, magtibay2023review}), depression (\cite{peng2019multivariate, he2022deep, lipschitz2019adoption}) and engagement (\cite{perkmann2021academic, saks2022organization, salam2022automatic, matamala2020role}), there is still a notable gap in understanding their interdependencies. None have simultaneously explored the methodologies, applications and challenges encompassing computational models for all three of these psychological states, leaving an opportunity to investigate how they intersect and influence one another. Stress has been examined from multiple perspectives, including its implications in workplace environments \cite{magtibay2023review}, its effects on drivers \cite{nvemcova2020multimodal} and its broader psychological impacts \cite{giannakakis2019review}. When examining depression, past surveys have delved into approaches using Electroencephalogram signals for deeper insights \cite{peng2019multivariate}, methods based on audio-visual information \cite{he2022deep} and the crucial connections between depression and patient engagement in healthcare settings \cite{lipschitz2019adoption}. Similarly, discussions on engagement have spanned diverse scenarios such as academic engagement \cite{perkmann2021academic}, engagement in workplace contexts \cite{saks2022organization}, inpatient care \cite{matamala2020role} and human-machine interactions \cite{salam2022automatic}. Despite the breadth of existing surveys on stress, depression and engagement, no unified work has yet examined all three together. This paper addresses that gap by presenting a comprehensive survey of computational approaches for analyzing these three psychological states in tandem, thereby providing a broader perspective on their collective significance.\vspace{.025in}

\begin{figure}[]
  \centering
  \hspace{-.12in}\includegraphics[width=.5\textwidth]{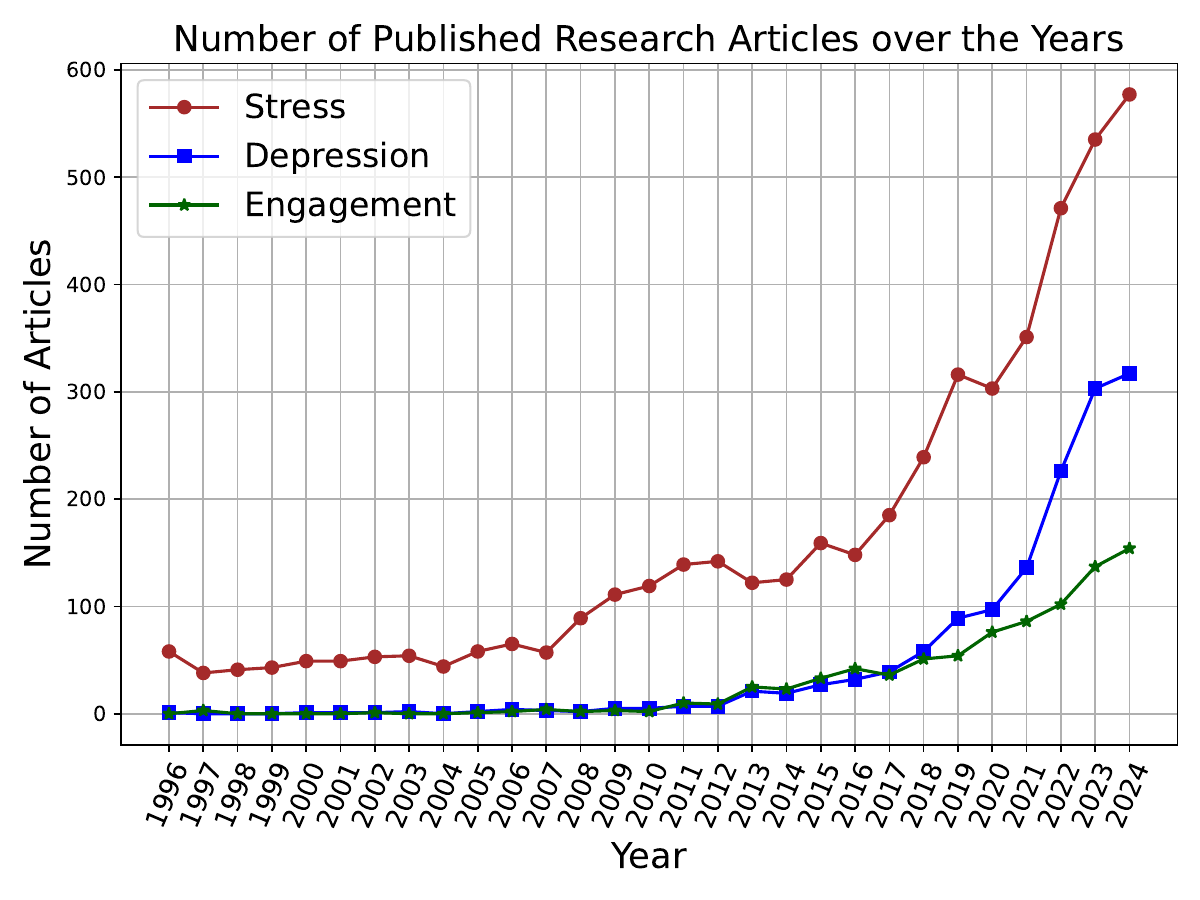}\vspace{-.1in}
  \caption{{Publication trends in stress, depression and engagement from 1996 to 2024 indicate a growing interest in these areas. This underscores the increasing importance of computational methods in addressing mental health challenges and enhancing well-being.}} \vspace{-.2in}
  \label{fig:trends}
\end{figure} 

{The computational approaches for analyzing these psychological states have seen substantial growth, as depicted in Fig. \ref{fig:trends}, with detailed trends and methodological evolutions discussed in Section \ref{sec:approaches}.} This paper surveys these approaches in a comprehensive manner, emphasizing their interconnectedness and the way advances in one area have influenced the others. For our literature review, we employed advanced queries in the Scopus database, targeting publications from 1994 to 2024 in top-tier journals and highly ranked conferences as per SCI and Qualis rankings. This search extended to include computational, psychological and neuroscience studies relevant to stress, depression and engagement, ensuring a broad and multi-disciplinary perspective. We manually reviewed the final list of papers to remove any irrelevant or redundant publications. Additionally, we made detailed notes on modalities, datasets used, contributions, methods, code availability, evaluation metrics, results and applications mentioned in these papers and used them during manuscript writing.\vspace{.025in} 

To aid the organization and flow of this survey paper, a detailed taxonomy outlining the computational analysis of stress, depression and engagement is presented in Table \ref{fig:taxonomy}. It summarizes various emotion categories, input modalities, computational approaches and applications of stress, depression and engagement analysis. The categorical and dimensional emotion classes are introduced in Section \ref{sec:intro}. Various datasets and input modalities for stress, depression and engagement analysis have been described in detail in Sections \ref{sec:datasets} and Section~\ref{sec:inputs} respectively. Section~\ref{sec:approaches} describes various computational approaches for stress, depression and engagement analysis along with their generic framework and state-of-the-art. The applications of these approaches have been discussed in Section \ref{sec:applications}. Section \ref{sec:futuredir} highlights challenges and future directions and Section \ref{sec:conc} concludes the paper.

\begin{table*}[!ht]
\centering
\caption{A taxonomy for computational analysis of stress, depression and engagement. The relevant research papers are presented based on emotion categories, input modalities (detailed in Section \ref{sec:inputs}), computational approaches (discussed in Section \ref{sec:sota}) and applications (mentioned in Section \ref{sec:applications}). Acronyms used include Machine Learning (ML), Deep Learning (DL), Convolutional Neural Network (CNN), Long Short Term Memory Network (LSTM), Gated Recurrent Unit (GRU), Support Vector Machine (SVM), K-Nearest Neighbour (KNN), Multimodal Learning (MM) and Bidirectional Encoder Representations from Transformers (BERT).\vspace{-.25in}}

\begin{tabularx}{\textwidth}{llll}
\multirow{56}{*}{\tikzmarknode{TaxonomyNode}{Taxonomy}} & & & \\
& \multirow{4}{*}{\tikzmarknode{EmotionsNode}{Emotions}} & & \\
&&\tikzmarknode{CategoricalNode}{Categorical}\parbox{13.3cm}{\cite{roseman1990appraisals,abedi2021affect, de2020encoding, chen2021sequential, li2022intelligent, xu2021privacy, niu2021hcag,  suparatpinyo2023smart, gupta2023facial, gupta2023multimodal, kastrati2023soaring, tao2023towards, salam2022automatic, savchenko2022classifying, mehta2022three, selim2022students, copur2022engagement, khenkar2022engagement, lee2022predicting}}&\vspace{.02in}\\  
&&\tikzmarknode{DimensionalNode}{Dimensional}\parbox{13.2cm}{\cite{barrett2017emotions, chen2023smg, carrizosa2021systematic, qi2023blockchain, delmastro2020cognitive, sun2022estimating, giannakakis2019review, rastgoo2018critical, can2019stress, roldan2021stressors, theerthagiri2023stress, nvemcova2020multimodal, turcan2019dreaddit,schmidt2018introducing,banerjee2023heart, naegelin2023interpretable}}&\\  

& \multirow{16}{*}{\tikzmarknode{ModalityNode}{\parbox{1cm}{\centering Inputs\\(Sec. \ref{sec:inputs})}}} & & \\
& & \multirow{5}{*}{\tikzmarknode{VisualNode}{Visual}}  
{\hspace{.2in}\tikzmarknode{Visual1Node}{Facial Features}}: \cite{he2018automatic,li2013spontaneous,gupta2023facial,chen2021sequential,selim2022students,liao2021deep,maddu2024online}&\vspace{.02in}\\
& & {\hspace{.5in}\tikzmarknode{Visual4Node}{Action Units}}: \cite{viegas2018towards,de2020encoding,alkabbany2019measuring,akbar2021exploiting}&\vspace{.05in}\\
& & {\hspace{.5in}\tikzmarknode{Visual3Node}{Eye Tracking Metrics}}: \cite{shen2022assessing, shan2020respiratory,savchenko2022classifying,choi2022immersion}&\vspace{.05in}\\
& & {\hspace{.5in}\tikzmarknode{Visual5Node}{Body Dynamics}}: \cite{kuttala2023multimodal,alghowinem2016multimodal,ashwin2020affective,chang2018}&\vspace{.07in}\\ 
& & {\hspace{.5in}\tikzmarknode{Visual2Node}{Micro-Gestures}}: \cite{chen2023smg, khenkar2022engagement, chen2019analyze, li2013spontaneous}&\vspace{.05in}\\
&& {\tikzmarknode{PhysioNode}{Physiological} 
\parbox{13cm}{\cite{carrizosa2021systematic,qi2023blockchain,delmastro2020cognitive,giannakakis2019review,rastgoo2018critical,can2019stress,roldan2021stressors,theerthagiri2023stress,schmidt2018introducing,banerjee2023heart,fernandez2018mental,zhu2023stress,xia2018physiological,saugbacs2020stress,zontone2019stress,ober2021detecting,peng2019multivariate,yu2022cloud, migovich2024stress, sawadogo2024ptsd, kerasiotis2024depression}}}&\vspace{.05in}\\ 
&& {\tikzmarknode{SpeechNode}{Audio}  \cite{huang2020domain,sardari2022audio,othmani2021towards,suparatpinyo2023smart,huang2018depression,abedi2021affect, talaat2024explainable, dogan2023multi}}&\vspace{.05in}\\ 
& & {\tikzmarknode{TextNode}{Text} \parbox{14cm}{\cite{turcan2019dreaddit,chiong2021textual,de2013predicting,zogan2023hierarchical,ragheb2021negatively,kastrati2023soaring,tao2023towards,de2021sadness,benini2019influence,li2022intelligent,xu2021privacy,kerasiotis2024depression, khowaja2024depression}}}\vspace{.05in}\\ 
& & {\tikzmarknode{MotionNode}{Motion} \parbox{14cm}
{\cite{liu2021learning, bobade2020stress, schmidt2018introducing}}}\vspace{.05in}\\ 
& & {\tikzmarknode{MultiNode1}{Multimodal}  \parbox{13.2cm}
{\cite{liao2021deep,zhu2020multi,alzoubi2012detecting, stappen2021muse,christ2022muse,naegelin2023interpretable,
niu2020multimodal,fang2023multimodal,casado2023depression,
rodrigues2019multimodal,alghowinem2016multimodal,li2023mha,zheng2023two,zhang2019multimodal,lin2020sensemood,niu2021hcag,celiktutan2017multimodal,arapakis2017interest,psaltis2017multimodal,ben2017ue, dogan2023multi, chen2024iifdd, xia2024depression, iyortsuun2024additive, tao2024depmstat, yang2023multimediate, sumer2021multimodal, li2024enhancing, dogan2023multi}}}\vspace{.015in}\\ 

& \multirow{18}{*}{\tikzmarknode{NetworkNode}{\parbox{1.1cm}{\centering Approaches\\(Sec. \ref{sec:approaches})}}} & & \\
& & \multirow{9}{*}{\hspace{.22in}\tikzmarknode{MLNode}{ML}} 
{\hspace{.2in}\tikzmarknode{ML1Node}{Logistic Regression}}: \cite{de2013predicting, migovich2024stress, awada2023new}&\vspace{.03in}\\
& & {\hspace{.62in}\tikzmarknode{ML2Node}{Decision Trees}}: \cite{carrizosa2021systematic, can2019stress, roldan2021stressors, dogan2023multi, mukhopadhyay2024tinystressnas}&\vspace{.03in}\\
& & {\hspace{.62in}\tikzmarknode{ML3Node}{Ensemble Methods}}: \cite{carrizosa2021systematic, roldan2021stressors, zhang2019multimodal, zhou2018visually, de2019depression, dogan2023multi, tanwar2024hybrid, khowaja2024depression}&\vspace{.03in}\\
& & {\hspace{.62in}\tikzmarknode{ML4Node}{Naive Bayes}}: \cite{roldan2021stressors, de2013predicting, awada2023new, dogan2023multi, sawadogo2024ptsd}&\vspace{.03in}\\
& & {\hspace{.62in}\tikzmarknode{ML5Node}{SVM and KNN}}: \cite{roldan2021stressors, ciman2016individuals, migovich2024stress, sawadogo2024ptsd}&\vspace{.03in}\\
& & {\hspace{.62in}\tikzmarknode{ML6Node}{Dimensionality Reduction}}: \cite{zhou2018visually, lin2020sensemood, de2020encoding, mukhopadhyay2024tinystressnas, li2024enhancing, lange2024generating}&\vspace{.03in}\\
& & {\hspace{.62in}\tikzmarknode{ML7Node}{Clustering}}: \cite{kastrati2023soaring, psaltis2017multimodal, arapakis2017interest}&\\&&& \vspace{-.05in}\\
& & \multirow{6}{*}{\hspace{.22in}\tikzmarknode{DLNode}{DL}} 
{\hspace{.2in}\tikzmarknode{DL1Node}{CNNs}}: \cite{zhou2018visually, he2021automatic, ts2020automatic, gupta2016daisee, quadrini2024stress, giakoumis2011automatic, fang2023multimodal, niu2020multimodal, casado2023depression, li2023mha, othmani2021towards, huang2020domain, dogan2023multi, tanwar2024hybrid}&\vspace{.03in}\\
& & {\hspace{.62in}\tikzmarknode{DL2Node}{RNN / LSTM / GRU}}: \cite{gupta2016daisee, casado2023depression, othmani2021towards, huang2020domain, talaat2024explainable}&\vspace{.03in}\\
& & {\hspace{.62in}\tikzmarknode{DL3Node}{ResNet}}: \cite{christ2022muse, chen2023smg, gupta2023facial, gupta2023multimodal, sawadogo2024ptsd}&\vspace{.03in}\\
& & {\hspace{.62in}\tikzmarknode{DL4Node}{Graph Neural Networks}}: \cite{adarsh2024mental, xu2024two, niu2021hcag, yang2023automatic, lu2024mast}&\vspace{.03in}\\
&& {\hspace{.62in}\tikzmarknode{DL5Node}{Attention Mechanism}}: \cite{de2020encoding, soares2022effects, fang2023multimodal, zheng2023two, lin2020sensemood, mallol2019hierarchical, chen2021sequential, tao2024depmstat}&\vspace{.03in}\\
& & {\hspace{.62in}\tikzmarknode{DL6Node}{Transformer / BERT}}: \cite{kastrati2023soaring, tao2023towards, kerasiotis2024depression, li2024enhancing, tao2024depmstat}&\vspace{.03in}\\
& & {\hspace{.62in}\tikzmarknode{DL7Node}{Autoencoder}}: \cite{sardari2022audio, niu2021hcag, xu2021privacy, khowaja2024depression, lange2024generating}&\\
& & {\hspace{.62in}\tikzmarknode{DL8Node}{Interpretable Deep Networks}}: \cite{zhou2018visually, chen2021sequential, tao2023towards, talaat2024explainable}&\vspace{.03in}\\
& & {\hspace{.62in}\tikzmarknode{DL9Node}{Federated Learning}}: \cite{li2022intelligent, xu2021privacy, lin2024wild}&\\
&&& \vspace{-.05in}\\


& & \multirow{2}{*}{\hspace{.22in}\tikzmarknode{DLMNode}{MM}} 
{\hspace{.175in}\tikzmarknode{MM1Node}{Multimodal Fusion}}: \cite{zheng2023two, casado2023depression, alghowinem2016multimodal, zhang2019multimodal, gupta2023multimodal,fang2023multimodal, niu2020multimodal, psaltis2017multimodal, zheng2023two, dogan2023multi, li2024enhancing}&\vspace{.04in}\\
& & {\hspace{.62in}\tikzmarknode{MM2Node}{Hybrid Models}}: \cite{ts2020automatic, gupta2016daisee, giakoumis2011automatic, fang2023multimodal, niu2020multimodal, casado2023depression, li2023mha, othmani2021towards, huang2020domain, tanwar2024hybrid}&\\
&&& \vspace{-.05in}\\

&&{\hspace{.22in}\tikzmarknode{A1Node}{Transfer Learning}}: \cite{theerthagiri2023stress,khenkar2022engagement,zhu2020multi,huang2020domain,suparatpinyo2023smart,mehta2022three,huang2020domain, ohse2024zero}&\vspace{.03in}\\

&&{\hspace{.2in}\tikzmarknode{A2Node}{Self-Supervised Learning}}: \cite{kuttala2023multimodal, fang2023multimodal, he2018automatic, yu2022cloud, zheng2023two, li2024enhancing}&\vspace{.03in}\\

&&{\hspace{.18in}\tikzmarknode{A3Node}{Human-Centered Computing}}: \cite{ashwin2020impact, ts2020automatic, zogan2023hierarchical, nvemcova2020multimodal, papadopoulos2016relative}&\vspace{.03in}\\

&&{\hspace{.12in}\tikzmarknode{A6Node}{Wearable Technologies}}: \cite{lin2020sensemood, delmastro2020cognitive, delmastro2020cognitive, anusha2019electrodermal, can2019stress, schmidt2018introducing, zhu2023stress, tanwar2024hybrid}&\vspace{.03in}\\

&&{\hspace{.16in}\tikzmarknode{A4Node}{Personalized Learning}}: \cite{gordon2016affective, can2019stress, ashwin2020impact, ts2020automatic}&\vspace{.03in}\\

&&{\hspace{.14in}\tikzmarknode{A5Node}{Blended Learning}}: \cite{salam2022automatic, sumer2021multimodal, magana2024ai, maddu2024online}&\vspace{.03in}\\

&&{\hspace{.1in}\tikzmarknode{A7Node}{Cloud-Edge Computing}}: \cite{yu2022cloud, chiang2022cognitive, mukhopadhyay2024tinystressnas, lin2024wild}&\\
&&& \vspace{-.1in}\\

& \multirow{16}{*}{\tikzmarknode{ApplicationsNode}{\parbox{1.33cm}{\centering Applications\\\hspace{.1in}(Sec. \ref{sec:applications})}}} & & \\
& & {\hspace{.1in}\tikzmarknode{Appl1Node}{Technological Solutions for Mental Health}: 
\cite{can2019stress, ciman2016individuals, kang2023k, schmidt2018introducing, anusha2019electrodermal, zhu2023stress, delmastro2020cognitive, xu2021privacy, migovich2024stress, sawadogo2024ptsd, kerasiotis2024depression, li2024enhancing}}&\vspace{.04in}\\

& & {\hspace{.1in}\tikzmarknode{Appl2Node}{Workplace and Occupational Well-being}:  
\parbox{9.83cm}{\cite{wiezer2013serious, rastgoo2018critical, nvemcova2020multimodal, mou2021driver, zontone2019stress,galanti2021work,travis2016m, upadyaya2016job, migovich2024stress, awada2023new}}}&\vspace{.03in}\\

& & {\hspace{.1in}\tikzmarknode{Appl3Node}{Detecting Mental Health Disorders}: 
\cite{zhang2019multimodal, ragheb2021negatively, o2023life, gurel2020automatic, kerasiotis2024depression, migovich2024stress, sawadogo2024ptsd, li2024enhancing}}&\vspace{.03in}\\

& & {\hspace{.1in}\tikzmarknode{Appl4Node}{Health and Behaviour Monitoring}: 
\cite{christ2022muse, nvemcova2020multimodal, jan2017artificial, li2022intelligent, niu2021hcag, gupta2023facial, casado2023depression, dogan2023multi}}&\vspace{.03in}\\

& & {\hspace{.1in}\tikzmarknode{Appl5Node}{Treatment Planning for Mental Health Disorders}: 
\parbox{8.85cm}{\cite{gupta2023multimodal, mallol2019hierarchical, banerjee2023heart, mou2021driver, zhu2023stress, li2023mha, delmastro2020cognitive, carrizosa2021systematic}}}&\vspace{.03in}\\

& & {\hspace{.1in}\tikzmarknode{Appl6Node}{Education and Learning Analytics}: 
\cite{selim2022students, shen2022assessing, sumer2021multimodal, gupta2023multimodal, lee2022predicting, ashwin2020impact, ashwin2020affective, ts2020automatic, kaur2018prediction, cho2017automated, lam2019context, maddu2024online, magana2024ai}}&\vspace{.03in}\\

& & {\hspace{.1in}\tikzmarknode{Appl7Node}{Gaming and Entertainment}: 
\cite{fang2023multimodal, niu2020multimodal, chen2019faceengage, ober2021detecting, giakoumis2011automatic, wiezer2013serious}}&\vspace{.03in}\\

& & {\hspace{.1in}\tikzmarknode{Appl8Node}{Human-Computer Interaction}:
\parbox{10.8cm}{\cite{gupta2023facial, gupta2023multimodal, savchenko2022classifying, shen2022assessing, abedi2021affect, ashwin2020affective, ts2020automatic, de2021sadness,  lin2020sensemood, psaltis2017multimodal, kang2023k, kastrati2023soaring, papadopoulos2016relative}}}&\vspace{.03in}\\

& & {\hspace{.1in}\tikzmarknode{Appl9Node}{Ethics and Privacy Preservation}:
\cite{de2013predicting, schmidt2018introducing, alghowinem2016multimodal, lam2019context, lange2024generating, lin2024wild}}&\vspace{.03in}\\

& & {\hspace{.1in}\tikzmarknode{Appl10Node}{Policy Making and Social Support}: 
\cite{ashwin2020affective, soares2022effects, kerasiotis2024depression, khowaja2024depression}}&\vspace{-.05in}\\
\end{tabularx} 
\tikz[remember picture,overlay]{
\draw[->,thick] (TaxonomyNode.east) -- ++(.15cm,0) -- ($(EmotionsNode.west) + (-0.25cm,0)$) -- (EmotionsNode.west);
\draw[->,thick] (TaxonomyNode.east) -- ++(.15cm,0) -- ($(ModalityNode.west) + (-.25cm,0)$) -- (ModalityNode.west);
\draw[->,thick] (TaxonomyNode.east) -- ++(.15cm,0) -- ($(NetworkNode.west) + (-.25cm,0)$) -- (NetworkNode.west);
\draw[->,thick] (TaxonomyNode.east) -- ++(.15cm,0) -- ($(ApplicationsNode.west) + (-.2cm,0)$) -- (ApplicationsNode.west);

\draw[->,thick] (EmotionsNode.east) -- ++(.15cm,0) -- ($(CategoricalNode.west) + (-0.25cm,0)$) -- (CategoricalNode.west);
\draw[->,thick] (EmotionsNode.east) -- ++(.15cm,0) -- ($(DimensionalNode.west) + (-0.25cm,0)$) -- (DimensionalNode.west);

\draw[->,thick] (ModalityNode.east) -- ++(.15cm,0) -- ($(PhysioNode.west) + (-0.25cm,0)$) -- (PhysioNode.west);
\draw[->,thick] (ModalityNode.east) -- ++(.15cm,0) -- ($(VisualNode.west) + (-0.25cm,0)$) -- (VisualNode.west);
\draw[->,thick] (VisualNode.east) -- ++(.15cm,0) -- ($(Visual1Node.west) + (-0.25cm,0)$) -- (Visual1Node.west);
\draw[->,thick] (VisualNode.east) -- ++(.15cm,0) -- ($(Visual2Node.west) + (-0.25cm,0)$) -- (Visual2Node);
\draw[->,thick] (VisualNode.east) -- ++(.15cm,0) -- ($(Visual3Node.west) + (-0.25cm,0)$) -- (Visual3Node.west);
\draw[->,thick] (VisualNode.east) -- ++(.15cm,0) -- ($(Visual4Node.west) + (-0.25cm,0)$) -- (Visual4Node.west);
\draw[->,thick] (VisualNode.east) -- ++(.15cm,0) -- ($(Visual5Node.west) + (-0.25cm,0)$) -- (Visual5Node.west);
\draw[->,thick] (ModalityNode.east) -- ++(.15cm,0) -- ($(SpeechNode.west) + (-0.25cm,0)$) -- (SpeechNode.west);
\draw[->,thick] (ModalityNode.east) -- ++(.15cm,0) -- ($(TextNode.west) + (-0.25cm,0)$) -- (TextNode.west);
\draw[->,thick] (ModalityNode.east) -- ++(.15cm,0) -- ($(MotionNode.west) + (-0.25cm,0)$) -- (MotionNode.west);
\draw[->,thick] (ModalityNode.east) -- ++(.15cm,0) -- ($(MultiNode1.west) + (-0.25cm,0)$) -- (MultiNode1.west);

\draw[->,thick] (NetworkNode.east) -- ++(.15cm,0) -- ($(MLNode.west) + (-0.25cm,0)$) -- (MLNode.west);
\draw[->,thick] (NetworkNode.east) -- ++(.15cm,0) -- ($(DLNode.west) + (-0.25cm,0)$) -- (DLNode.west);
\draw[->,thick] (NetworkNode.east) -- ++(.15cm,0) -- ($(DLMNode.west) + (-0.25cm,0)$) -- (DLMNode.west);
\draw[->,thick] (NetworkNode.east) -- ++(.15cm,0) -- ($(A1Node.west) + (-0.25cm,0)$) -- (A1Node.west);
\draw[->,thick] (NetworkNode.east) -- ++(.15cm,0) -- ($(A2Node.west) + (-0.25cm,0)$) -- (A2Node.west);
\draw[->,thick] (NetworkNode.east) -- ++(.15cm,0) -- ($(A3Node.west) + (-0.25cm,0)$) -- (A3Node.west);
\draw[->,thick] (NetworkNode.east) -- ++(.15cm,0) -- ($(A4Node.west) + (-0.25cm,0)$) -- (A4Node.west);
\draw[->,thick] (NetworkNode.east) -- ++(.15cm,0) -- ($(A5Node.west) + (-0.25cm,0)$) -- (A5Node.west);
\draw[->,thick] (NetworkNode.east) -- ++(.15cm,0) -- ($(A6Node.west) + (-0.25cm,0)$) -- (A6Node.west);
\draw[->,thick] (NetworkNode.east) -- ++(.15cm,0) -- ($(A7Node.west) + (-0.25cm,0)$) -- (A7Node.west);

\draw[->,thick] (MLNode.east) -- ++(.15cm,0) -- ($(ML1Node.west) + (-0.25cm,0)$) -- (ML1Node.west);
\draw[->,thick] (MLNode.east) -- ++(.15cm,0) -- ($(ML2Node.west) + (-0.25cm,0)$) -- (ML2Node.west);
\draw[->,thick] (MLNode.east) -- ++(.15cm,0) -- ($(ML3Node.west) + (-0.25cm,0)$) -- (ML3Node.west);
\draw[->,thick] (MLNode.east) -- ++(.15cm,0) -- ($(ML4Node.west) + (-0.25cm,0)$) -- (ML4Node.west);
\draw[->,thick] (MLNode.east) -- ++(.15cm,0) -- ($(ML5Node.west) + (-0.25cm,0)$) -- (ML5Node.west);
\draw[->,thick] (MLNode.east) -- ++(.15cm,0) -- ($(ML6Node.west) + (-0.25cm,0)$) -- (ML6Node.west);
\draw[->,thick] (MLNode.east) -- ++(.15cm,0) -- ($(ML7Node.west) + (-0.25cm,0)$) -- (ML7Node.west);

\draw[->,thick] (DLNode.east) -- ++(.15cm,0) -- ($(DL1Node.west) + (-0.25cm,0)$) -- (DL1Node.west);
\draw[->,thick] (DLNode.east) -- ++(.15cm,0) -- ($(DL2Node.west) + (-0.25cm,0)$) -- (DL2Node.west);
\draw[->,thick] (DLNode.east) -- ++(.15cm,0) -- ($(DL3Node.west) + (-0.25cm,0)$) -- (DL3Node.west);
\draw[->,thick] (DLNode.east) -- ++(.15cm,0) -- ($(DL4Node.west) + (-0.25cm,0)$) -- (DL4Node.west);
\draw[->,thick] (DLNode.east) -- ++(.15cm,0) -- ($(DL5Node.west) + (-0.25cm,0)$) -- (DL5Node.west);
\draw[->,thick] (DLNode.east) -- ++(.15cm,0) -- ($(DL6Node.west) + (-0.25cm,0)$) -- (DL6Node.west);
\draw[->,thick] (DLNode.east) -- ++(.15cm,0) -- ($(DL7Node.west) + (-0.25cm,0)$) -- (DL7Node.west);
\draw[->,thick] (DLNode.east) -- ++(.15cm,0) -- ($(DL8Node.west) + (-0.25cm,0)$) -- (DL8Node.west);
\draw[->,thick] (DLNode.east) -- ++(.15cm,0) -- ($(DL9Node.west) + (-0.25cm,0)$) -- (DL9Node.west);


\draw[->,thick] (DLMNode.east) -- ++(.15cm,0) -- ($(MM1Node.west) + (-0.25cm,0)$) -- (MM1Node.west);
\draw[->,thick] (DLMNode.east) -- ++(.15cm,0) -- ($(MM2Node.west) + (-0.25cm,0)$) -- (MM2Node.west);

\draw[->,thick] (ApplicationsNode.east) -- ++(.15cm,0) -- ($(Appl1Node.west) + (-0.25cm,0)$) -- (Appl1Node.west);
\draw[->,thick] (ApplicationsNode.east) -- ++(.15cm,0) -- ($(Appl2Node.west) + (-0.25cm,0)$) -- (Appl2Node.west);
\draw[->,thick] (ApplicationsNode.east) -- ++(.15cm,0) -- ($(Appl3Node.west) + (-0.25cm,0)$) -- (Appl3Node.west);
\draw[->,thick] (ApplicationsNode.east) -- ++(.15cm,0) -- ($(Appl4Node.west) + (-0.25cm,0)$) -- (Appl4Node.west);
\draw[->,thick] (ApplicationsNode.east) -- ++(.15cm,0) -- ($(Appl5Node.west) + (-0.25cm,0)$) -- (Appl5Node.west);
\draw[->,thick] (ApplicationsNode.east) -- ++(.15cm,0) -- ($(Appl6Node.west) + (-0.25cm,0)$) -- (Appl6Node.west);
\draw[->,thick] (ApplicationsNode.east) -- ++(.15cm,0) -- ($(Appl7Node.west) + (-0.25cm,0)$) -- (Appl7Node.west);
\draw[->,thick] (ApplicationsNode.east) -- ++(.15cm,0) -- ($(Appl8Node.west) + (-0.25cm,0)$) -- (Appl8Node.west);
\draw[->,thick] (ApplicationsNode.east) -- ++(.15cm,0) -- ($(Appl9Node.west) + (-0.25cm,0)$) -- (Appl9Node.west);
\draw[->,thick] (ApplicationsNode.east) -- ++(.15cm,0) -- ($(Appl10Node.west) + (-0.25cm,0)$) -- (Appl10Node.west);
}
\label{fig:taxonomy}\vspace{-.1in}
\end{table*}

\section{{Datasets}}\label{sec:datasets}
\begin{figure*}[]
  \centering
  \includegraphics[width=.96\textwidth]{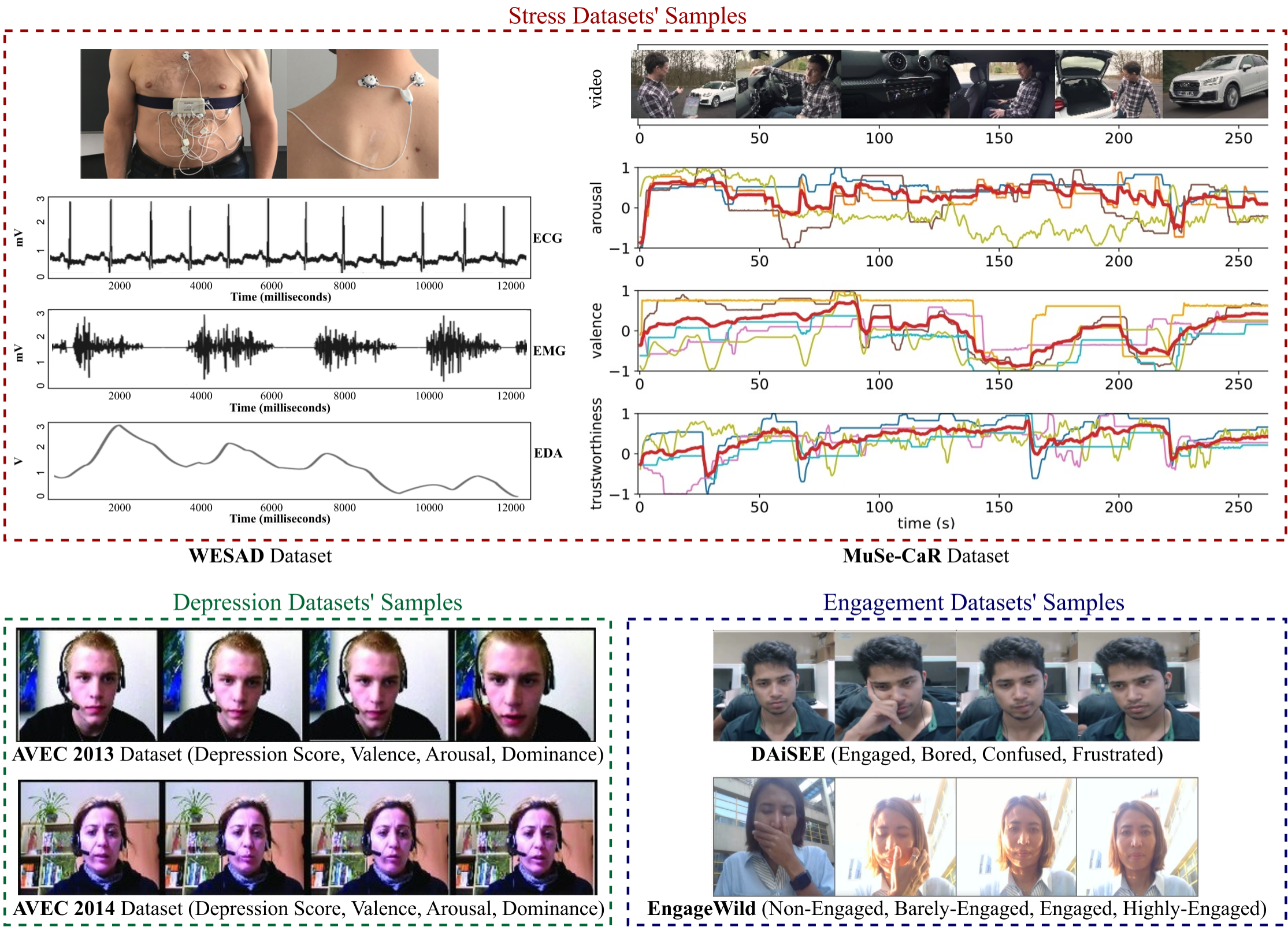} \vspace{-.09in}
  \caption{Sample data inputs and modalities for most commonly used datasets mentioned in Section \ref{sec:sota_results}. Along with audio-visual modalities, they use physiological signals such as ECG, EDA and EMG and have labels for valence, arousal, dominance, trustworthiness, depression and engagement categories.}\vspace{-.12in}
  \label{fig:inputs}
\end{figure*}

Table \ref{tab:dataset_summary} summarizes stress, depression and engagement datasets, {with sizes given in hours except for unimodal text-only datasets where duration is not applicable.} Their details are discussed in the following sections and sample inputs are shown in Fig. \ref{fig:inputs}.

\begin{table*}
\centering
\caption{Summary of datasets for Stress, Depression and Engagement analysis. Here `A', `M', `P', `T' and `V' represent `Audio', `Motion', `Physiological', `Textual' and `Visual' modalities, respectively and
`\textit{GT}' specifies the type of Ground Truth label used, with `HA' for Humanly Annotated, `TD' for Task Determined, `CA' for Clinically Assessed and `SR' for Self Report.\vspace{-.07in}}

\label{tab:dataset_summary}
\resizebox{1\textwidth}{!}
{
    \begin{tabular}{lllclcc}
    \toprule
    \textbf{Name} & \textbf{Year} & \textbf{Focus Area} & \textbf{Size (hours)} & \textbf{Subjects} & \textbf{Modalities}  & \textbf{GT} \\
    \midrule
    \multicolumn{7}{c}{\textbf{Stress}} \\
    \midrule
    \href{https://project.inria.fr/stressid/}{StressID} \cite{chaptoukaev2023stressid} & 2023  & Multiple Stimuli Stress & 39  & 65 Adults & APV  & SR \\
    \href{https://github.com/mikecheninoulu/SMG}{SMG} \cite{chen2023smg} & 2023 & Micro-Gestures & 8.14 & 40 Adults & V & HA\\
    \href{https://ieee-dataport.org/open-access/maus-dataset-mental-workload-assessment-n-back-task-using-wearable-sensor}{MAUS} \cite{beh2021maus} & 2021 & Workload Stress & 12.83 & 22 Adults & P & TD+SR \\
    \href{https://sites.google.com/view/muse-2021}{ULM-TSST} \cite{stappen2021muse} & 2021 & Public Stress & 5.78 & 105 Participants & P & SR \\
    \href{https://sites.google.com/view/muse-2021/challenge/data}{MuSe-CaR} \cite{stappen2021multimodal} & 2021 & Driver Stress & 40.2 & 38 Drivers & APTV  & HA \\
    \href{https://hubbs.engr.tamu.edu/resources/verbio-dataset/}{VerBIO} \cite{yadav2020exploring} & 2021 & Bio-behaviour & 180 & 55 Students & APT & SR \\
    \href{https://ieee-dataport.org/open-access/database-cognitive-load-affect-and-stress-recognition}{CLAS} \cite{markova2019clas} & 2019 & Workplace Stress & 31 & 62 Adults & APT  & TD+SR\\ 
    \href{https://ieee-dataport.org/open-access/dasps-database}{DASPS} \cite{baghdadi2019dasps} & 2019 & Anxiety Analysis & 1.15 & 23 Participants & P  & SR\\  
    \href{https://ubicomp.eti.uni-siegen.de/home/datasets/icmi18/}{WESAD} \cite{schmidt2018introducing} & 2018 & Wearable Stress & 13.38 & 15 Adults & PM & SR\\
    \href{https://zenodo.org/record/7086222}{Passau-SFCH} \cite{christ2022multimodal} & 2017 & Stress in Sports & 11 & 10 Footballers &ATV & HA \\
    \href{https://archive.physionet.org/physiobank/database/drivedb/}{DRIVEDB} \cite{healey2000smartcar} & 2000 & Driver Stress & 21.97 & 17 Drivers  & P & ED\\
    \bottomrule 
    \multicolumn{7}{c}{\textbf{Depression}} \\
    \midrule
    
    \href{https://hacilab.github.io/MPDDChallenge.github.io/}{MPDD} \cite{mpdd2025}  & 2025 & Depression \& Personality & 9.68 & 228 Participants  & ATV  & CA \\ 
    \href{https://ieee-dataport.org/open-access/chinese-multimodal-depression-corpus}{CMDC} \cite{zou2022semi} & 2023 & Semi-structured Interviews & 1.41 & 167 Adults & ATV 
    & CA \\
    \href{https://github.com/zzzzzzyang/MMDA-a-Multimodal-Dataset-for-Depression-and-Anxiety-Detection}{MMDA} \cite{jiang2022mmda} & 2022 & Clinical Interviews & 48.05 & 1025 Participants & ATV & CA \\
    \href{https://github.com/Fancy-Block/EATD-Corpus}{EATD} \cite{shen2022eatd} & 2022 & Short Q\&A Interviews & 2.26 & 162 Participants & AT & SR \\
    \href{https://sites.google.com/view/jeewoo-yoon/dataset}{D-Vlog} \cite{yoon2022dvlog} & 2022 & Vlog Recordings & 160 & 816 Participants & AV & HA \\
    \href{https://modma.lzu.edu.cn/data/index/}{MODMA} \cite{cai2022modma} & 2020 & Mental Disorder Analysis & 49.5 & 55 Participants & AP & CA \\
    \href{https://researchoutput.ncku.edu.tw/en/publications/data-collection-of-elicited-facial-expressions-and-speech-respons}{Chi-Mei} \cite{huang2018detecting} & 2020 & Mood Database & 6.8 & 11 Participants & A & CA \\ 
    \href{https://dcapswoz.ict.usc.edu/}{Extended DAIC} \cite{ringeval2019avec} & 2019 & Clinical Interviews & 71.38 & 185 Adults & ATV & SR\\
    \href{https://www.researchgate.net/profile/Zhaocheng-Huang-3/publication/327389040_Depression_Detection_from_Short_Utterances_via_Diverse_Smartphones_in_Natural_Environmental_Conditions/links/5b9345604585153a5305d798/Depression-Detection-from-Short-Utterances-via-Diverse-Smartphones-in-Natural-Environmental-Conditions.pdf}{SH2} \cite{huang2018depression} & 2018 & Phone Ctterances & 16 & 887 Participants & A & TD+CA\\
    \href{http://users.umiacs.umd.edu/~resnik/umd_reddit_suicidality_dataset.html}{UM Suicidality} \cite{shing2018expert} & 2018 & Suicidality Reports & \textendash & 934 Subjects & T &  HA+CA\\ 
    \href{https://erisk.irlab.org/}{eRisk} \cite{losada2019overview} & 2017 & Health Risk Detection & \textendash & 4427 Subjects & T & HA+SR\\ 
    \href{https://dcapswoz.ict.usc.edu/}{DAIC-WOZ} \cite{gratch2014distress} & 2016 & Clinical Interviews & 50.21 & 189 Adults & ATV & SR \\
    \href{https://www.blackdoginstitute.org.au/resources-support/depression/}{BlackDog} \cite{alghowinem2012joyous} & 2016 & Open Ended Questions & 8.55 & 130 Participants & A & CA  \\
    \href{https://www.cs.jhu.edu/~mdredze/clpsych-2015-shared-task-evaluation/}{CLPsych} \cite{coppersmith2015clpsych} & 2015 & Mental Health Posts & \textendash & 1989 Subjects & T & HA+SR\\ 
    \href{http://avec2013-db.sspnet.eu/}{AVEC 2014} \cite{valstar2014avec} & 2014 & Emotion Challenge & 4.52 & 58 Adults & AV & SR\\
    \href{http://avec2014-db.sspnet.eu/}{AVEC 2013} \cite{valstar2013avec} & 2013 & Emotion Challenge & 39.5 & 58 Adults & AV & SR \\
    \href{https://diuf.unifr.ch/main/diva/recola/}{RECOLA} \cite{ringeval2013introducing} & 2013 & Remote Collaboration & 9.5 & 46 Participants & APV & HA+SR\\ 
    \href{https://www.researchgate.net/publication/256309258_Detecting_Depression_Severity_from_Vocal_Prosody}{Pittsburgh} \cite{yang2012detecting} & 2012 & Clinical Interviews & 3.61 & 49 Participants & A & CA \\ 
    \bottomrule
    \multicolumn{7}{c}{\textbf{Engagement}} \\
    \midrule
    \href{https://sites.google.com/view/dreams-dataset}{DREAMS} \cite{singh2024dreams} & 2024 & Engagement \& Attention & 8.68 & 32 Adults & V & SR \\
    \href{https://sites.google.com/view/emotiw2023/home?authuser=0}{EngageNet}\cite{singh2023engagenet} & 2023 & Student Engagement & 31 & 127 Adults & V & HA+SR\\
    \href{https://nmsl.kaist.ac.kr/projects/attention/}{PAFE}\cite{lee2022predicting}  & 2022 & Student Engagement & 15 & 15 Adults & V & SR\\
    \href{https://ieeexplore.ieee.org/document/9893134}{VRESEE} \cite{selim2022students} & 2022 & Student Engagement & 9.79 & 88 Adults &  V  & HA+SR\\ 
    \href{http://sh.rice.edu/project/face-engage/}{FaceEngage} \cite{chen2019faceengage} & 2019 & Gameplay Engagement & 2.18 & 25 Adults & V  & HA\\
    \href{https://sites.google.com/view/emotiw2018}{EngageWild} \cite{kaur2018prediction} & 2018 & Student Engagement & 16.5 & 91 Adults & V  & HA\\    
    \href{https://adasp.telecom-paris.fr/resources/2017-05-18-ue-hri/}{UE-HRI} \cite{ben2017ue} & 2017 & Human-Robot Interaction & 24.98 & 54 Adults & AV & HA \\
    \href{https://www.cl.cam.ac.uk/research/rainbow/projects/mhhri/}{MHHRI} \cite{celiktutan2017multimodal} & 2017 & Human-Robot Interaction & 6 & 18 Adults & APV  & SR \\
    \href{https://vcl.iti.gr/dataset/masr-dataset/}{MASRD} \cite{psaltis2017multimodal} & 2017 & Games for Students & 0.625 & 15 Subjects & V  & HA \\ 
    \href{https://people.iith.ac.in/vineethnb/resources/daisee/index.html}{DAiSEE} \cite{gupta2016daisee} & 2016 & Student Engagement & 25 & 112 Adults  & V &  HA \\ \bottomrule    
    \end{tabular}
} \vspace{-.1in}
\end{table*}

\subsection{Datasets for Stress Analysis}
\subsubsection{Unimodal Datasets} 
Spontaneous Micro-Gesture (SMG) dataset \cite{chen2023smg} presents data on stress and micro-gesture, comprising 821056 frames (8 hours of video) from 40 adults. Mental workload Assessment on n-back task Using wearable Sensor (MAUS) dataset \cite{beh2021maus}, collected from 22 adults and Ulm Trier Social Stress Test (TSST) dataset \cite{stappen2021muse}, with data from 105 participants gathered, focus on stress and mental workload. Stress Recognition in automobile Driver database (DRIVEDB) \cite{healey2000smartcar}, released in 2000 by Healey and Picard at MIT's Media Lab, pioneered drivers' stress detection with physiological data from 17 drivers in a lab setting and expert-derived Ground Truth labels. DASPS dataset \cite{baghdadi2019dasps} contains physiological sensor data from 23 participants to evaluate anxiety.

\subsubsection{Multimodal Datasets} 
The \textit{StressID} dataset \cite{chaptoukaev2023stressid} focuses on multiple stimuli offering 39 hours of audio-physiological-visual data from 65 adults. \textit{MuSe-CAR} dataset includes 303 recordings from 38 drivers captured using audio-visual and physiological sensors, providing insights into drivers' stress in real-world scenarios. \textit{CLAS} dataset \cite{markova2019clas} focuses on occupational stress, comprising 31 hours of audio and physiological data from 62 adults, while the \textit{WESAD} dataset \cite{schmidt2018introducing}, collected using questionnaires from 15 adults, contains physiological and motion modality information. \textit{VerBIO} dataset \cite{yadav2020exploring} delves into bio-behavioral stress with 180 hours of data from 55 students, collected using audio, physiological and thermal sensors across various settings. \textit{Passau-SFCH} dataset \cite{christ2022multimodal} explores sports-related stress, comprising 11 hours of audio, video and thermal data, capturing insights from footballers in environment.

\subsection{Datasets for Depression Analysis} 
\subsubsection{Unimodal Datasets} 
\textit{Sonde Health Free Speech (SH2-FS)} dataset \cite{huang2018depression} includes recordings of individuals in everyday settings such as cars, homes and workplaces. Its annotations draw upon the Patient Health Questionnaire (PHQ) self-diagnostic test \cite{kroenke2001phq}. Further, the \textit{eRisk} dataset looks at health risks using over a million sentences from 4427 people. The \textit{CLPsych} dataset \cite{coppersmith2015clpsych} was constructed using text posts by 1989 subjects while the \textit{Pittsburgh} dataset \cite{yang2012detecting} was collected utilizing audio recordings from 49 people. Similarly, the \textit{BlackDog} dataset \cite{alghowinem2012joyous} includes audio recordings from 130 participants in open-ended question-answer format. The \textit{Chi-Mei Mood disorder database} \cite{huang2018detecting} from the Chi-Mei Medical Center focuses on mood disorders. The \textit{UM (University of Maryland)} dataset \cite{shing2018expert} focuses on depression within an academic context.

\subsubsection{Multimodal Datasets} 
The 2013 \cite{valstar2013avec} and 2014 \cite{valstar2014avec} variants of \textit{Audio Visual Emotion Challenge (AVEC)} pioneered the datasets for depression analysis using audio-visual indicators. Subsequently, the \textit{Distress Analysis Interview Corpus - Wizard of Oz (DAIC-WOZ)} dataset \cite{gratch2014distress} was introduced in 2014 and further enhanced to construct the \textit{Extended DAIC} dataset in 2019 \cite{ringeval2019avec}, both focusing on clinical interviews to assess depression. The \textit{RECOLA} multimodal database \cite{ringeval2013introducing}, developed through interdisciplinary collaboration at Université de Fribourg, integrates physiological, audio and video signals for remote depression assessment. The \textit{MPDD} dataset \cite{mpdd2025} (2025) expands depression research with 228 sessions of audio, textual and visual data. Similarly, the \textit{CMDC} \cite{zou2022semi} and \textit{MMDA} \cite{jiang2022mmda} datasets capture depression indicators from 167 and 1025 clinical interview sessions, respectively. The \textit{EATD} dataset \cite{shen2022eatd} focuses on 486 short Q\&A interview audios for speech-based screening, while \textit{D-Vlog} \cite{yoon2022dvlog} analyzes 961 vlog recordings from 816 individuals to examine depression markers. The \textit{MODMA} dataset \cite{cai2022modma} contains EEG and audio information from 55 participants, linking neural and vocal features for depression detection.

\subsection{Datasets for Engagement Analysis}
\subsubsection{Unimodal Datasets} 
The \textit{DAiSEE} dataset \cite{gupta2016daisee} provides 9068 video clips from 112 users, capturing boredom and engagement in e-learning, while \textit{EngageWild} \cite{kaur2018prediction} presents 264 videos from 91 subjects across four engagement levels. Expanding on this, \textit{EngageNet} \cite{singh2023engagenet} includes 31 hours of data from 127 participants with over 11,300 clips, exploring both behavioral and cognitive engagement. Similarly, \textit{PAFE} \cite{lee2022predicting} features 15 hours of videos and 1,100 attention probes from 15 students, analyzing attention shifts in online lectures. The \textit{VRESEE} dataset \cite{selim2022students} extends engagement analysis to 88 Egyptian students with 3,525 recorded videos. Beyond educational contexts, \textit{FaceEngage} \cite{chen2019faceengage} contains over 700 YouTube gaming videos from 25 amateur gamers, assessing engagement variations based on demographics and gameplay duration. The \textit{MASRD} dataset \cite{psaltis2017multimodal} further integrates immersion and presence using the Game Engagement Questionnaire (GEQ). More recently, the \textit{DREAMS} dataset \cite{singh2024dreams} introduced 8.7 hours of video from 32 participants, capturing engagement and attention across diverse real-world scenarios.

\subsubsection{Multimodal Datasets} 
The \textit{Multimodal Human-Human-Robot Interaction (MHHRI)} dataset \cite{celiktutan2017multimodal} analyses interactions between humans and robots. It offers engagement and personality metrics from 18 participants during dyadic and triadic interactions. The \textit{User Engagement in Human-Robot Interaction (UE-HRI)} dataset \cite{ben2017ue} documents spontaneous human interactions with the robot Pepper, focusing on facial expressions and postural details related to engagement.

\subsection{Discussion of Datasets}
The aforementioned datasets are crucial for analyzing stress, depression and engagement, yet challenges arise due to differences in modalities, data sizes and labeling techniques. Multimodal datasets like \textit{MuSe-CaR}, \textit{VerBIO}, \textit{DAIC}, \textit{RECOLA} and \textit{MHHRI} incorporate diverse data types, making synchronization difficult \cite{chen2019analyze}. Addressing these challenges requires standardized formats and protocols to enable seamless cross-dataset comparisons and integrative analyses. {Notably, to the best of our knowledge, no publicly available dataset jointly annotates any two or all three of these psychological states (stress, depression, engagement), highlighting a key gap in affective computing research.} Future datasets should use multi-label annotations to capture overlapping states and refine engagement modeling by distinguishing passive and active engagement, especially in stress or depression contexts. 

\section{Inputs}\label{sec:inputs}
Various input modalities used in analyzing stress, depression and engagement are outlined below. 

\subsection{Unimodal Inputs}
\subsubsection{Visual Modality}\label{sec:visual-modality}
The following specific clues are often extracted from images and videos for analysing stress, depression and engagement. \vspace{.04in}

\begin{itemize}[itemsep=1pt]
\item \textit{Facial Features}: Landmark coordinates, textures and expressions represent physical characteristics, especially in the eyes, nose and mouth regions. They are analyzed to identify emotions like stress and engagement through facial images and videos. He et al. \cite{he2018automatic} used dynamic facial appearance for stress analysis, while Li et al. \cite{li2013spontaneous} applied representation learning for depression recognition. In another work, Gupta et al. \cite{gupta2023facial} used facial emotion recognition in real-time online settings for engagement detection.

\item \textit{Action Units (AUs)}: AUs describe facial muscle movements linked to emotions and can detect subtle states like stress or suppressed emotions \cite{viegas2018towards}. To this end, De et al. \cite{de2020encoding} explored AU encoding for stress, depression and engagement analysis while Alkabbany et al. \cite{alkabbany2019measuring} measured engagement by analyzing AUs during learning activities.

\item \textit{Eye Tracking Metrics}: Gaze, blink rate and saccadic movements help understand focus and attention. They correlate with cognitive engagement \cite{shen2022assessing, shan2020respiratory}. In this direction, Savchenko et al. \cite{savchenko2022classifying} and Choi et al. \cite{choi2022immersion} analyzed eye tracking for attention during online learning and video watching.

\item \textit{Body Dynamics}: Body dynamics, including head movements, posture and body language, convey emotional states through non-verbal cues. For example, Kuttala et al. \cite{kuttala2023multimodal} explored body dynamics for stress detection, while Alghowinem et al. \cite{alghowinem2016multimodal} used head posture, movements and eye gaze for depression detection.

\item \textit{Micro-Gestures}: Micro-gestures are subtle, involuntary movements reflecting inner feelings \cite{chen2019analyze}. In this context, Chen et al. \cite{chen2023smg} studied their interplay with emotion states and utilized them for emotional stress analysis. 
\end{itemize}

\subsubsection{Physiological Modality}\label{sec:physiological-modality}
The frequently used methods to measure physiological responses for stress, depression and engagement analysis are described below. They provide insights about special physiological characteristics associated with stress, depression and engagement \cite{alzoubi2012detecting}.  

\begin{itemize}[itemsep=1pt]
\item \textit{Heart Rate Activity}: It includes monitoring heart rate variability (HRV) and patterns using Electrocardiogram (ECG) and Photoplethysmogram (PPG) sensors. It is particularly useful for detecting stress and engagement levels, as changes in heart rate can indicate emotional arousal or relaxation. For instance, Giannakakis et al. \cite{giannakakis2019stress} utilized HRV for stress analysis. Additionally, remote PPG, which can be classified under both physiological and visual modalities, has been employed in works like those of Sun et al \cite{sun2022estimating} and Casado et al. \cite{casado2023depression}.

\item \textit{Electroencephalogram (EEG)}: EEG measures brain activity and is often used in the analysis of depression and stress. It provides insight into cognitive processing and emotional regulation. In this context, Xia et al. \cite{xia2018physiological} and Sharma et al. \cite{sharma2022evolutionary} used EEG for stress and depression analysis, demonstrating its effectiveness in detecting nuanced changes in mental states.

\item \textit{Electrodermal Activity (EDA)}: EDA, also known as skin conductance, measures the electrical changes on the skin surface due to sweat gland activity, which is influenced by the sympathetic nervous system. It is a sensitive marker for emotional arousal, stress and engagement. For example, {while Zhu et al. \cite{zhu2023stress} used EDA to detect stress, Alzoubi et al. \cite{alzoubi2012detecting} reported EDA-based features that help in identifying both depression-related and engagement-related arousal changes.} 

\item \textit{Electromyogram (EMG)}: EMG measures the electrical activity produced by skeletal muscles and is indicative of muscle tension, often related to stress or emotional intensity. It's used for analyzing facial muscle responses in emotional states and stress. In this direction, Pourmohammadi et al. \cite{pourmohammadi2020stress} exemplify the use of EMG along with other biosignals for stress detection.
 
\item \textit{Respiratory Signals}: Respiratory rate and breathing patterns are critical indicators of psychological states like stress or relaxation. Changes in breathing can reflect emotional arousal, stress, or engagement levels. For instance, respiratory signals have been used by Shan et al. \cite{shan2020respiratory} and Fernandez et al. \cite{fernandez2018mental} for stress detection.
\end{itemize}

\subsubsection{Audio Modality}\label{sec:speech-modality}
{Techniques like prosodic analysis, voice quality and speech rate extract features from audio signals to detect stress, depression and engagement. Huang et al. \cite{huang2020domain} used domain adaptation in speech emotion recognition, while Sardari et al. \cite{sardari2022audio} emphasized audio features for predicting emotions. Suparatpinyo et al. \cite{suparatpinyo2023smart} leveraged acoustic features for stress and depression and Chen et al. \cite{chen2021sequential} studied intonation and loudness for emotion recognition. Ben et al. \cite{ben2017ue} used audio analysis to monitor engagement with the robot Pepper, showing speech cues capture engagement levels. Rejaibi et al. \cite{rejaibi2022mfcc} employed MFCC-based RNN for depression recognition, underscoring speech analysis's clinical relevance.}

\subsubsection{Text Modality}\label{sec:text-modality}
{Textual data from social media and online platforms also play a pivotal role in identifying stress, depression and engagement. Turcan et al. \cite{turcan2019dreaddit} examined Reddit discussions to detect stress-related expressions, while Chiong et al. \cite{chiong2021textual} proposed methods for depressive symptom detection. In educational contexts, Kastrati et al. \cite{kastrati2023soaring} showed that text analytics could gauge student engagement by examining discourse cues in online forums. Othmani et al. \cite{othmani2021towards} employed linguistic features for affect and depression recognition and Lin et al. \cite{lin2020sensemood} introduced SenseMood for depression detection on social media. Together, these works underscore text-based approaches’ adaptability across the three psychological states, offering insight into users’ mental well-being.}

\subsubsection{Motion Modality}\label{sec:motion-modality}
{Motion plays a crucial role in understanding and interpreting physical responses to psychological stressors. It is used to analyze bodily movements as indicators of stress levels. Notably, Schmidt et al. \cite{schmidt2018introducing} incorporated motion data to establish a foundational approach for stress analysis. Following this, Bobade et al. \cite{bobade2020stress} enhanced the integration of motion with physiological signals, showing significant improvements in model accuracy and robustness. Similarly, Liu et al. \cite{liu2021learning} employed a client-server model that leveraged motion data to refine multimodal representations for stress detection, emphasizing the critical nature of motion data in comprehensive behavioural analyses.} 

\subsection{{Multimodal Inputs}}
Real-world analyses of stress, depression and engagement often combine multiple signals such as visual, textual, audio and physiological, for a holistic perspective. For instance:

\begin{itemize}[itemsep=1pt] 
\item \textit{Audio+Visual (A+V)}: Merging speech features like pitch, with facial or gestural cues like micro-expressions uncovers subtle indicators of stress, depression, or engagement \cite{celiktutan2017multimodal}.

\item \textit{Audio+Text (A+T)}: Adding text to audio improves depression and stress detection, especially with negative self-focus \cite{orabi2018deep}.

\item \textit{Visual+Physiological (V+P)}: Pairing facial video with biometrics enhances stress and engagement analysis by linking expression to arousal \cite{mou2021driver}.

\item \textit{Physiological+Audio (P+A)}: Linking biosignals like ECG and EDA to vocal prosody helps validate emotional arousal and detect stress in real-time \cite{rastgoo2018critical}.

\item \textit{Physiological+Motion (P+M)}: Pairing wearable sensor data with accelerometer readings helps capture dynamic changes in stress and engagement \cite{schmidt2018introducing}.

\item \textit{Audio+Visual+Text (A+V+T)}: Incorporating lingual features with audio-visual content enables a holistic analysis by fusing semantics, intonation and appearance \cite{celiktutan2017multimodal}.

\item \textit{Audio+Visual+Physiological (A+V+P)}: Integrating speech prosody, facial cues and biometric measures enables robust mood disorder detection and engagement monitoring \cite{celiktutan2017multimodal}.
\end{itemize}

\subsection{Discussion of Inputs} 
Analyzing stress, depression and engagement is challenging due to their complex nature compared to typical emotions \cite{ekman1992argument, carrizosa2021systematic}. Input cues vary by task: subtle micro-gestures \cite{chen2019analyze} serve as effective visual indicators for high-stress detection, particularly when individuals mask their feelings \cite{qi2023blockchain, sun2022estimating}. Physiological signals capture prolonged, subtle changes in stress, while visual modalities detect rapid fluctuations \cite{carrizosa2021systematic,chen2019analyze}. Audio and textual data provide additional emotional context not captured by other modalities \cite{huang2018depression,chiong2021textual}. Multimodal approaches, combining these inputs, represent the state-of-the-art for analyzing these states \cite{alghowinem2016multimodal, celiktutan2017multimodal}. Unlike typical emotions with clear cues (e.g., smiling for `happy’), these states vary significantly across individuals, making multimodal methods essential for improved understanding and performance \cite{kumar2023interpretable}. To effectively capture this variability, selecting fusion techniques tailored to the data’s unique characteristics is crucial for enhancing the accuracy and robustness of computational models \cite{sanchez2023predictive, kuttala2023multimodal}.

 
\section{Computational Approaches for Stress, Depression and Engagement Analysis}\label{sec:approaches}
This section discusses how computational methods analyze stress, depression and engagement by using data from digital interactions, wearables and sensors to create accurate models. These methods, preferred over traditional ones relying on subjective reports and clinical observations, offer a more objective and effective understanding of complex emotional states \cite{fang2023multimodal, giannakakis2019review}.

\begin{figure*}[]
  \centering
  \includegraphics[width=1\textwidth]{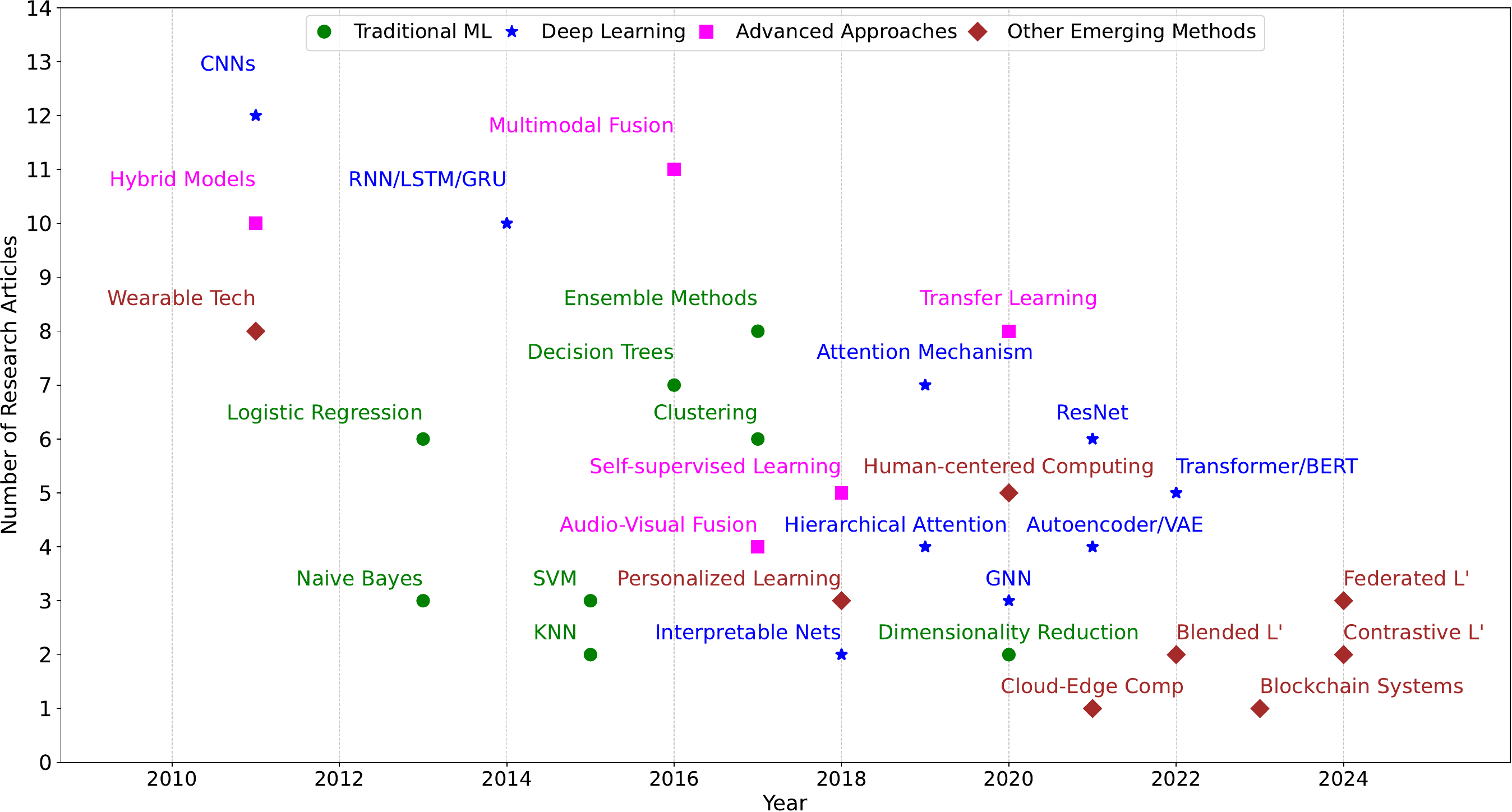}\vspace{-.07in}
  \caption{{Chronological emergence of computational approaches from traditional ML techniques (e.g., Naive Bayes, Logistic Regression) to advanced DL methods (e.g., GNNs, Transformers) and emerging paradigms (e.g., federated learning, cloud-edge computing), illustrating the evolving complexity and sophistication of stress, depression and engagement analysis. Here, L' denotes `Learning'.}}
  \label{fig:timeline}\vspace{-.12in}
\end{figure*}

\subsection{ {Categories of Computational Approaches}}
Figure~\ref{fig:timeline} outlines the evolution of computational approaches for analyzing stress, depression and engagement, showing the shift from basic to advanced learning strategies. This section categorizes these approaches into three groups: Machine Learning, Deep Learning and Advanced Learning Approaches. We discuss these paradigms: Supervised learning, using fully labeled data; Unsupervised learning, operating without explicit labels (e.g., anomaly detection); Semi-supervised learning, utilizing a mix of labeled and unlabeled data; and Self-supervised learning, where models derive training signals from the data itself, enhancing efficiency \cite{vedernikov2024analyzing, sun2024, sarkar2020self}.

\subsubsection{Traditional Machine Learning}\label{sec:ml-approach}
Machine Learning (ML) approaches empower computers to learn from data without requiring explicit rule-based instructions. They often involve hand-engineered features combined with relatively simpler classifiers, particularly in \textit{supervised} settings. Historically, ML predominated before deep learning gained traction and it remains a strong choice for certain tasks where data availability is limited or model interpretability is crucial \cite{zhu2023stress, naegelin2023interpretable}. 

In stress analysis, early ML solutions incorporate physiological signals or smartphone sensors to build classical models. For instance, Saugbacs et al.~\cite{saugbacs2020stress} relied on accelerometer and gyroscope data to train decision tree and KNN classifiers for stress detection. Likewise, in depression research, De et al.~\cite{de2013predicting} introduced a logistic regression framework to diagnose major depressive disorder from social media posts. Classifiers like Naive Bayes, ensemble techniques and random forests have also been employed for multi-modal, small-scale tasks, including depressive state recognition \cite{zhang2019multimodal}. 

Beyond fully supervised pipelines, traditional ML encompasses \textit{unsupervised} approaches—such as clustering to group stress or depression indicators—and \textit{semi-supervised} methods, where partially labeled data guides the learning process. Roldan et al.~\cite{roldan2021stressors}, for example, used anomaly detection to flag outlier stress signals in a scenario with minimal labels. Moreover, ML-based engagement analysis typically relies on interpretable features (e.g., facial action units, gaze metrics) to gauge attention and immersion. Taken together, these traditional ML methods still excel when computational resources are constrained, datasets are small, or transparency in decision-making is paramount \cite{zhu2023stress, naegelin2023interpretable}.

\subsubsection{Deep Learning}\label{sec:dl-approach}
Deep Learning (DL) is a subset of ML employing neural architectures with multiple layers to automatically learn representations from raw data. Its popularity rose around 2010 (Fig.~\ref{fig:timeline}). Unlike traditional ML, DL often reduces reliance on hand-crafted features~\cite{he2021automatic}. Basic \textit{supervised DL} techniques such as CNNs and RNNs are extensively applied to single-modality data like facial images or text. Zhou et al.~\cite{zhou2018visually} used CNNs for depression detection from facial images. Zhong et al.~\cite{zhong2023robust} leveraged CNN-based discriminant features for robust depression classification. In the textual domain, Orabi et al.~\cite{orabi2018deep} and Cai et al.~\cite{cai2023depression} applied neural networks for depression recognition on social media. Other language-oriented works have used RNNs, BERT and Transformers to analyze depression \cite{zhu2020multi}, highlighting the growing role of large-scale language models. These neural methods achieve strong performance in stress recognition~\cite{gupta2023facial, quadrini2024stress} and engagement prediction~\cite{iyortsuun2024additive, mehta2022three}.

Recent innovations in DL use \textit{unsupervised, semi-Supervised and self-Supervised} approaches. In this context, Hierarchical attention networks, graph neural networks and Transformers expand beyond simple CNN/RNN architectures~\cite{mallol2019hierarchical, tanwar2024hybrid, adarsh2024mental}. Fang et al.~\cite{fang2023multimodal} used multi-level attention with limited labeled data (a semi-supervised scenario) to detect depression. Kuttala et al.~\cite{kuttala2023multimodal} integrated self-supervised hierarchical CNNs for stress analysis. Sun et al.~\cite{sun2024} introduced an unsupervised/weakly supervised remote physiological measurement approach using spatiotemporal contrast. He et al.~\cite{he2018automatic} and Yu et al.~\cite{yu2022cloud} leveraged weak supervision to detect depression from facial data with minimal labels. In addition, context-aware systems and voice source analysis have also been explored to personalize emotion recognition~\cite{lam2019context}. These strategies reduce annotation costs and often yield robust solutions when labeled data are expensive or scarce.

\subsubsection{Advanced Approaches}
Many approaches beyond classical ML or DL used in stress, depression and engagement analysis are discussed below.

\begin{itemize}[itemsep=1pt]
\item \textbf{Transfer Learning.} Transfer Learning reuses models or features from one task or dataset for a new, possibly related, task. This is highly beneficial in mental health contexts. In stress detection, Theerthagiri et al.~\cite{theerthagiri2023stress} and Albaladejo et al.~\cite{albaladejo2023evaluating} improved classification accuracy by adapting pre-trained networks to new stress datasets. For depression, domain adaptation or advanced voice-recognition models are repurposed, as in~\cite{huang2020domain, suparatpinyo2023smart}. Engagement analysis has similarly seen successful transfer, for example, Khenkar et al.~\cite{khenkar2022engagement} leveraged micro-gesture embeddings to interpret learner behavior. Such supervised or semi-supervised transfer solutions accelerate model convergence, handle smaller labeled sets and facilitate domain generalization.

\item \textbf{Multimodal Learning and Fusion Methods.}
Multimodal learning integrates two or more different data streams (e.g., visual, audio, text, physiological) to obtain a richer representation of psychological states. Chen et al.~\cite{chen2024iifdd} showed the benefit of combining intra- and inter-modal features from IoMT data for depression detection. Xia et al.~\cite{xia2024depression} employed a multimodal graph neural network for structured signals in depression detection, while He et al.~\cite{he2022deep} demonstrated how integrating EEG, speech and facial expressions can improve recognition accuracy. Mou et al.~\cite{mou2021driver} merged physiological signals with driver behavior for stress detection and Orabi et al.~\cite{orabi2018deep} combined text plus acoustic data for depression analysis. To systematically fuse these diverse inputs, following fusion strategies are employed.

\begin{itemize}[itemsep=1pt]
\item \textit{Early Fusion (Feature-Level Fusion)}: Features from multiple modalities are merged before classification to costruct a comprehensive feature vector that encompasses information from all the modalities. It uses techniques like concatenation, stacking and weighted fusion \cite{sanchez2023predictive}. 

\item \textit{Late Fusion (Decision-Level Fusion)}: Each modality is processed and classified independently. The resulting classifications or scores from each modality are then fused to make a final decision. It is beneficial when each modality provides strong, independent evidence for the classification. Methods like ensemble learning, majority voting and score-level fusion are commonly used. \cite{mou2021driver}.
\end{itemize}

\end{itemize}

\subsubsection{Other Emerging Approaches}  
Emerging methods are addressing the complexities of real-world mental health analysis through innovative paradigms. \textit{Federated Learning} \cite{li2022intelligent, xu2021privacy} decentralizes training to preserve privacy, a critical feature for handling sensitive medical data. \textit{Contrastive Learning} \cite{li2024enhancing} enhances robustness by learning representations through comparisons of positive and negative sample pairs. \textit{Cloud-Edge Computing} \cite{yu2022cloud} optimizes real-time detection of depression and stress by balancing local (on-device) and remote computation. \textit{Personalized Learning} \cite{gordon2016affective, ts2020automatic, ashwin2020impact} tailors detection models to individual characteristics, improving accuracy. \textit{Blended Learning} \cite{salam2022automatic} integrates traditional and online education, fostering flexible environments that support well-being. \textit{Human-Centered Computing and Wearable Technologies} are also advancing the field: Chen et al. \cite{chen2023smg} utilized body gestures for stress analysis, Shan et al. \cite{shan2020respiratory} developed non-contact respiratory sensors and Lin et al. \cite{lin2020sensemood} combined wearable signals with social media data for depression monitoring. Schmidt et al. \cite{schmidt2018introducing} constructed the WESAD dataset for multimodal wearable stress and affect detection. Additionally, secure frameworks like blockchain-based systems \cite{qi2023blockchain} are being explored, highlighting the growing emphasis on robust, privacy-preserving solutions.
 
\subsubsection{Discussion of Categories of Approaches} 
The field of stress, depression and engagement analysis has seen significant expansion since 2010 with the adoption of machine ML techniques, further accelerated by advanced DL approaches over the following decade \cite{giannakakis2019review}. This evolution from traditional methods towards ML and DL has not only improved analytical performance \cite{arapakis2017interest} but also shifted the computational landscape towards automated data representation extraction, reducing reliance on manual feature engineering \cite{he2021automatic}. Despite these advances, DL technologies come with challenges, including high computational demands, the need for large labeled datasets and complexities in model interpretability \cite{de2013predicting, roldan2021stressors}. Recently, the emphasis on interpretable deep networks has grown, highlighting the importance of transparency and trust in mental health applications \cite{li2022intelligent}. Moreover, hybrid and multimodal fusion techniques have become popular, balancing performance with computational efficiency \cite{gupta2023facial}.

\subsection{Computational Analysis Framework}\label{sec:sota} 
\subsubsection{Generic Phases}

\begin{figure*}[!h]
  \centering
  \includegraphics[width=1\textwidth]{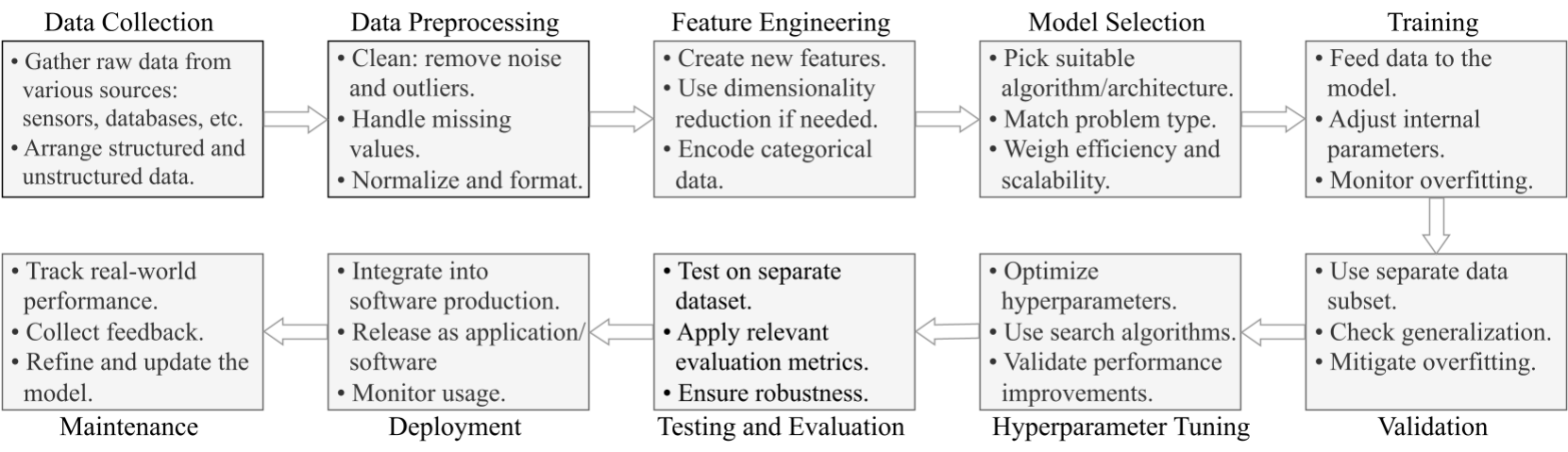}\vspace{-.12in}
  \caption{A depiction of the generic phases used in various computational approaches for stress, depression and engagement analysis.}\vspace{-.1in}
  \label{fig:pipeline}
\end{figure*}
 
A general framework for constructing computational models for stress, depression and engagement analysis is depicted in Fig. \ref{fig:pipeline}. Initially, data acquisition and preprocessing eliminate noise and extraneous information. Subsequent transformations involve feature engineering, dimensionality reduction and encoding. The models undergo training and evaluation with advanced methods like K-fold cross-validation \cite{pampouchidou2020automated} and hyperparameter tuning via grid and random search to enhance performance \cite{yu2022cloud}. In industrial settings, deployment integrates these models into applications, refining predictions with user feedback. These deployment phases are less emphasized in academic research, which instead prioritizes advancements in concepts, methodologies and algorithms.

\subsubsection{Data Collection and Preprocessing}\label{sec:prepro}
Data for stress, depression and engagement analysis collected from sources such as video, audio, wearable sensors and online platforms undergo preprocessing before further analysis.\vspace{.02in}

\begin{itemize}[itemsep=1pt]
\item \textit{Visual Data Preprocessing}: 
This involves face detection and tracking across video frames which is crucial for dynamic emotion analysis \cite{giannakakis2019review}. Facial landmark detection is key for precise emotion recognition, pinpointing critical facial expressions \cite{stappen2021muse}. Noise removal techniques like histogram equalization are also employed for visual data preprocessing \cite{casado2023depression}. The face registration and alignment methods are used to standardize the data for consistent facial analysis \cite{christ2022muse}. 

\item \textit{Physiological Signal Preprocessing}:
Physiological signals like HRV and EDA provide insight into stress or emotional states. Noise removal in these signals is vital for accurate analysis, typically involving filtering methods to eliminate irrelevant artefacts \cite{zhang2019multimodal}. The Z-score-based normalization is also used to improve the comparability across subjects \cite{banerjee2023heart}. Further, dimensionality reduction techniques like PCA help focus on relevant signal aspects while reducing complexity \cite{naegelin2023interpretable}.

\item \textit{Speech Preprocessing}: 
In speech analysis, extracting features that reflect stress or other emotional states is crucial. Features such as pitch, energy and formant frequencies capture key emotional cues. Normalization (e.g., min-max scaling) adjusts for variations in loudness and speaking rate \cite{rejaibi2022mfcc}. Speech noise removal filters out background disturbances for clearer feature extraction \cite{talaat2024explainable}. Dimensionality reduction methods like MFCCs distill the speech features.

\item \textit{Text Preprocessing}:
Textual data analysis involves processing raw text to extract meaningful patterns. Tokenization, stop-word removal and stemming/lemmatization break down the text into analyzable elements \cite{roldan2021stressors}. Normalization like lowercasing and punctuation removal ensures uniformity. Techniques like TF-IDF or word embeddings reduce dimensionality, capturing the text's semantic essence \cite{turcan2019dreaddit}. 

\item {\textit{Motion Preprocessing}: This step involves noise filtering and sensor calibration for accuracy, alongside principal component analysis to reduce dimensionality and focus on essential motion features \cite{liu2021learning, bobade2020stress}. Time synchronization and normalization align and scale the data, facilitating its integration with other modalities for comprehensive analysis \cite{schmidt2018introducing, deldari2022cocoa}.}\vspace{-.03in}

\end{itemize}

\subsubsection{Model Selection and Training Strategies}\label{sec:training}
Training computational models for stress, depression and engagement analysis presents several challenges \cite{he2021automatic}. These include handling small and biased datasets \cite{theerthagiri2023stress}, ensuring model robustness \cite{chen2023smg}, efficiently using multimodal data \cite{sharma2022evolutionary} and preventing overfitting during model training \cite{ragheb2021negatively}. To address these challenges, the following strategies are employed:\vspace{.02in}

\begin{itemize}[itemsep=1pt]
\item \textit{Handling Small and Biased Data Sets}: Addressing the issues of small and biased datasets is crucial for creating reliable models. Data augmentation methods help by expanding the variety and amount of training data, which reduces overfitting risks \cite{zhao2024driver}. Similarly, Transfer Learning proves beneficial by applying knowledge from related tasks, enhancing model performance with limited data \cite{wu2019multi}.

\item \textit{Ensuring Model Robustness}: Transfer learning and ensemble models boost model robustness and accuracy. Transfer learning applies insights from related tasks to improve learning, while ensemble methods merge multiple models' predictions, diversifying decision-making and enhancing emotion and stress recognition \cite{othmani2021towards}.

\item \textit{Utilizing Attention Mechanism for Time-Series Data}: The attention mechanism enables the models to focus on crucial sections of time-series data selectively. This technique is particularly beneficial for accurately analyzing stress by examining physiological and behavioural signals, providing a more nuanced understanding of these indicators \cite{he2021automatic}.

\item \textit{Integrating Multimodal Data for Comprehensive Analysis}: Integrating data from various sources, including ECG, EEG and wearable sensors, is crucial for creating detailed models. This approach to fuse complementary information from multiple modalities improves emotion analysis accuracy \cite{sharma2022evolutionary}.

\item \textit{Adopting Enhanced Learning Strategies and Feature Fusion}: Using adaptive learning rates and structured feature fusion enhances training efficiency with multi-source data, improving learning speed, consistency and pattern recognition across physiological signals \cite{kuttala2023multimodal}.
\end{itemize}

\subsubsection{Testing and Evaluation Techniques}\label{sec:eval}
Models for analyzing stress, depression and engagement are assessed through following methods. 

\begin{itemize}[itemsep=1pt]   
    \item \textit{Quantitative Evaluation}:
    This approach uses numerical metrics such as Accuracy (Acc), Mean Absolute Error (MAE), Root Mean Squared Error (RMSE), precision, recall and F1-score to evaluate model performance. Models are automatically evaluated against standard benchmarks for both general and specific analysis, removing the need for manual assessment \cite{beltran2019analysis}. 
    
    \item \textit{Qualitative Evaluation}: 
    In qualitative assessment, the focus is on descriptively analyzing the model's output to gauge its capture of emotional context. This sheds light on capabilities and areas for enhancement beyond what quantitative measures reveal. For instance, Chen et al. \cite{chen2023deep} analyzed a stress detection model's reaction to video types and Lin et al. \cite{lin2022deep} studied the impact of voice quality on depression recognition, linking pronunciation to predictions.
    
    \item \textit{Human Evaluation}:
    Experts such as psychologists or clinicians conduct evaluations, applying both quantitative and qualitative measures based on their understanding of emotional nuances \cite{gupta2016daisee}. Although this method is more demanding in terms of time and resources compared to automated assessments, it provides an in-depth evaluation, which is especially valuable for identifying stress, depression and engagement states.
\end{itemize}

\subsection{{State-of-the-Art}}\label{sec:sota_results}
While Section \ref{sec:sota} presents a generic framework for computational analysis of stress, depression and engagement, this section discusses the latest advancements for the same.

\subsubsection{State-of-the-Art for Stress Analysis}
Stress detection research has expanded rapidly, particularly on the WESAD dataset, where newer methods integrate advanced data augmentation and deep networks to surpass 95\% accuracy. For example, Li et al. \cite{li2025multimodal} employ ConvNeXt and GAN-based augmentation on physiological signals, while Wang et al. \cite{wang2024physioformer} introduce PhysioFormer, achieving 99.54\% accuracy through specialized feature extraction. Self-supervised and contrastive techniques are also on the rise: Pulse-PPG \cite{saha2025pulse} targets PPG signals with over 94\% accuracy and COCOA \cite{deldari2022cocoa} contrasts multiple sensor modalities. Federated approaches \cite{liu2021learning} further secure data privacy by distributing training across client devices without centralizing raw samples. Beyond wearable-only solutions, MuSe-CaR \cite{stappen2021multimodal} focuses on driver stress with multimodal inputs (audio-visual, textual), while Siamese Capsule methods \cite{kothuri2024siamese} handle continuous stress prediction. This landscape underscores a shift toward more robust, privacy-preserving techniques that fuse multiple signals and leverage powerful augmentation or self-supervision for high-precision stress detection. Table \ref{tab:performance_summary_stress} presents a comprehensive performance evaluation of state-of-the-art methods for stress analysis.

\subsubsection{State-of-the-Art for Depression Analysis}
As discussed in Table \ref{tab:performance_summary_depression}, depression analysis increasingly leverages the AVEC benchmarks, where multimodal techniques capture subtle affective and physiological cues. Dictionary-based methods \cite{niu2025depression} apply bidirectional fusion across audio, visual and textual data, reducing mean absolute errors below 4.0. Spatiotemporal fusion \cite{wang2025automatic} extends these gains by examining continuous facial and physiological patterns. Time-domain speech modeling \cite{li2025efficient} refines acoustic features through dual-path attention, boosting detection robustness. In parallel, unsupervised rPPG-based analysis \cite{casado2023depression} uncovers remote physiological changes via face videos, yielding a less intrusive approach. Advanced CNN architectures (e.g., MMDepNet \cite{kumar2025wacv}) combine multiple data streams such as physiological, textual and visual to enhance feature granularity. Collectively, these works show a trend toward deeper neural architectures, multimodal synergy and domain-adaptive learning, effectively lowering errors and improving real-world viability in both clinical and everyday contexts.

\subsubsection{State-of-the-Art for Engagement Analysis}
Engagement detection, though less studied than stress and engagement, is gaining traction on datasets like EngageWild \cite{kaur2018prediction} and DAiSEE \cite{gupta2016daisee}. In EngageWild, real-time CNN solutions \cite{savchenko2022classifying} utilize EfficientNet or MobileNet variants for on-device inference, while multi-segment LSTM and TCN approaches \cite{huynh2019engagement, abedi2021affect} enhance frame-level capture of student affect. On DAiSEE, EngageFormer \cite{mandia2025transformer} integrates physiological and visual data through a transformer design and self-supervised ViT-based autoencoders \cite{zhang2024self} tackle facial feature reconstruction for higher accuracy. Other solutions incorporate spatiotemporal networks, like DFSTN \cite{liao2021deep}, or ordinal classification \cite{abedi2021improving} to manage fine-grained engagement levels. Overall, the field is moving toward multimodal fusion and deeper architectures, leveraging flexible sequences or hybrid attention blocks to handle complex behaviors across diverse learning scenarios. Table \ref{tab:performance_summary_engagement} presents a comprehensive performance evaluation of state-of-the-art methods for engagement analysis.


\begin{table*}[!h]
\centering
\caption{Performance summary of stress analysis approaches, sorted first by year and then by Accuracy (`Acc'). Here `CCC,' `V,' `P,' `A,' and `M' denote Concordance Correlation Coefficient, visual, physiological, audio and motion modalities. The abbreviations used include COCOA (Cross Modality Contrastive Learning for Sensor Data), PCA (Principal Component Analysis), ANN (Artificial Neural Network), RF (Random Forest), DT (Decision Tree), GAN (Generative Adversarial Network) and SVM (Support Vector Machine).\vspace{-.07in}} 
\label{tab:performance_summary_stress}
\resizebox{.96\textwidth}{!}
{
    \begin{tabular}{|C{.8cm}|p{4.8cm}|C{.8cm}|p{4.7cm}|C{.9cm}|C{.8cm}|C{.8cm}|}\toprule
    \textbf{Dataset} & \textbf{Method} & \textbf{Year} & \textbf{Architecture} & \textbf{Modality} & \textbf{CCC} & \textbf{Acc}\\ \midrule
    \multicolumn{6}{c}{\textbf{Stress}} \\ \midrule   
    
    \multirow{13}{*}{\rotatebox{90}{\href{https://ubicomp.eti.uni-siegen.de/home/datasets/icmi18/}{WESAD}\cite{schmidt2018introducing}}}
    & Wearables' feature analysis \cite{alkurdi2025extending} & 2025 & XGBoost + DT + Transfer learning & P & \textendash & 99.00\% \\
    & Pulse-PPG \cite{saha2025pulse} & 2025 & Contrastive learning-based model & P & \textendash & 94.52\% \\
    & PhysioFormer \cite{wang2024physioformer} & 2024 & ContribNet + AffectNet & P & \textendash & 99.54\% \\
    & Data augmentation \cite{li2025multimodal} & 2024 & ConvNeXt + Self-attention GAN & P & \textendash & 95.70\% \\
    & Self-supervised learning \cite{wu2023transformer} & 2023 & Temporal convolution + Transformer & P & \textendash & 96.29\% \\
    & Cross-modality contrastive learning \cite{deldari2022cocoa} & 2022 & COCOA & PM & \textendash & 97.60\% \\
    & Multimodal representation \cite{liu2021learning} & 2021 & Client-server aggregated model & PM & \textendash & 93.20\% \\ 
    & Self-supervised learning \cite{sarkar2020self} & 2020 & Multi-task CNN & P  & \textendash & 96.90\% \\
    & Multimodal bio-signal analysis \cite{bobade2020stress} & 2020 & PCA + ANN + RF & PM & \textendash & 95.21\% \\
    & Multimodal fusion \cite{bota2020emotion} & 2020 & Bimodal Deep AutoEncoder & P & \textendash & 90.25\% \\
    & Base paper (two-class) \cite{schmidt2018introducing} & 2018 & SVM + ANN + RF & PM & \textendash & 93.12\% \\
    & Multimodal bio-signals (three-class) \cite{bobade2020stress} & 2018 & PCA + ANN + RF & PM & \textendash & 84.32\%\\
    & Base paper (three-class) \cite{schmidt2018introducing} & 2018 & SVM + ANN + RF & PM & \textendash & 80.34\% \\
    \hdashline

    \multirow{8}{*}{\rotatebox{90}{\href{https://sites.google.com/view/muse-2021/challenge/data}{MuSe-CaR}\cite{stappen2021multimodal}}}
    & SCapsNet \cite{kothuri2024siamese} & 2024 & Siamese + Capsule Net + Optimization & AVT & 0.5072 & \textendash\\
    & DeepSpectrum \cite{christ2022muse} & 2022 & Early fusion + Attention & APVT & 0.4585 & \textendash\\
    & Multitask learning \cite{jiang2021multitask} & 2021 & Self-attention + Bi-LSTM + EfficientNet & AV & 0.3587 & \textendash\\
    & Attention enhanced recurrent net \cite{sun2021multimodal} & 2021 & VGGFace + DeBERTa + Attention & ATV & 0.3803 & \textendash\\
    & Hybrid (early + late) fusion \cite{christ2022muse} & 2021 &  VGGish + VGGface + OpenFace + BERT & APVT & 0.4646 & \textendash \\ 
    & Aligned Annotation Weighting \cite{stappen2021muse} & 2021 & LSTM + RNN & APVT & 0.4913 & \textendash \\ 
    & Multimodal sentiment analysis \cite{stappen2021multimodal} & 2021 & BERT + FastText + VGGish & ATV & 0.5384 & \textendash\\
    & Attention enhanced recurrent net \cite{sun2021multimodal} & 2021 & Wav2vec + DeBERT + Attention & AV & 0.5558 & \textendash\\
    

    \bottomrule
    \end{tabular}
}
\end{table*}

\begin{table*}[!h]
\centering
\caption{Performance summary of depression analysis approaches, sorted first by year and then by mean absolute error (`MAE'). Here, `RMSE,' `V,' `P,' `A,' and `T' denote root mean square error, visual, physiological, audio and textual modalities. The acronyms used include SSD (Single Shot Multibox Detection network), RFR (Random Forest Regressor), LGBPTOP (Local dynamic appearance descriptor), LPQ (Local Phase Quantisation), C3D/3DCNN (3D CNN), 2DCNN (2D CNN), FDHH (Feature Dynamic History Histogram), SVR (Support Vector Regression), LPQ (Local Phase Quantization) and TOP (Temporal Occurrence Pattern).\vspace{-.07in}}  
\label{tab:performance_summary_depression}
\resizebox{.96\textwidth}{!}
{
    \begin{tabular}{|C{.75cm}|p{4.8cm}|C{.8cm}|p{4.8cm}|C{.9cm}|C{.8cm}|C{.8cm}|}\toprule
    \textbf{Dataset} & \textbf{Method} & \textbf{Year} & \textbf{Architecture} & \textbf{Modality} & \textbf{MAE} & \textbf{RMSE}\\ \midrule 
    \multirow{10}{*}{\rotatebox{90}{\href{http://avec2013-db.sspnet.eu/}{AVEC 2013} \cite{valstar2013avec}}} 
    & Dictionary‐based decomposition \cite{niu2025depression} & 2025 & Bidirectional multimodal fusion & ATV & 3.87 & 5.21 \\
    & Long-term spatio-temporal routing \cite{wang2025automatic} & 2025 & Spatiotemporal fusion ensemble network & PV & 5.38 & 6.74 \\
    & Time-domain speech modeling \cite{li2025efficient} & 2025 & Dual-path state-space attention network & A & 8.35 & 9.05 \\ 
    & Multimodal Depression analysis \cite{kumar2025wacv} & 2024 & MMDepNet & APTV & 6.15 & 7.89 \\ 
    & Unsupervised rPPG-based analysis \cite{casado2023depression} & 2023 &SSD + LGBPTOP + RFR + ResNet-50 & PV & 6.43 & 8.01\\
    & Depth-wise convolution analysis \cite{niu2021multi} & 2021 & 3DCNN + SVR & AV & 6.19 &  8.02 \\ 
    & Two-stream image analysis \cite{de2020encoding} & 2020 & Two-stream 2DCNN & V & 5.96 &  7.97 \\ 
    & Deep residual learning \cite{de2019depression} & 2019 & ResNet-50 & V & 6.30 &  8.25 \\ 
    & Multi-channel ensembling \cite{zhou2018visually} & 2018 & Four DCNNs & V & 6.20 &  8.28 \\ 
    & Depth-wise video analysis \cite{al2018video} & 2018 & C3D & V & 7.37 & 9.28 \\ 
    & Dual-channel analysis \cite{zhu2017automated} & 2017 & Two DCNN & V & 7.58 &  9.82 \\ 
    & Local pattern analysis \cite{wen2015automated} & 2015 & LPQ-TOP + MFA & V & 8.22 & 10.27 \\ 
    & Eye-based feature extraction \cite{kaya2014eyes} & 2014 & LPQ + Geo & AV & 7.86 &  9.72 \\ 
    & Baseline paper \cite{valstar2013avec} & 2013 & OpenSMILE + LGBP-TOP & AV & 10.88 & 13.61 \\ 
    \hdashline
    \multirow{9}{*}{\rotatebox{90}{\href{http://avec2014-db.sspnet.eu/}{AVEC 2014} \cite{valstar2014avec}}} 
    &  Dictionary‐based decomposition \cite{niu2025depression} & 2025 & Bidirectional multimodal fusion & ATV & 3.63 & 5.05 \\
    & Long-term spatio-temporal routing \cite{wang2025automatic} & 2025 & Spatiotemporal fusion ensemble network & PV & 5.09 & 6.83 \\
    & Time-domain speech modeling \cite{li2025efficient} & 2025 & Dual-path state-space attention network & A & 8.39 & 9.14  \\ 
    & Multimodal Depression Analysis \cite{kumar2025wacv} & 2024 & MMDepNet & APTV & 6.14 & 8.11 \\ 
    & Unsupervised rPPG-based analysis \cite{casado2023depression} & 2023 & SSD + LGBPTOP + RFR + ResNet-50 & PV & 6.57 & 8.49\\ 
    & Depth-wise convolutional analysis \cite{niu2021multi} & 2021 & 3DCNN + SVR & AV & 6.14 &  7.98 \\ 
    & Two-stream image analysis \cite{de2020encoding} & 2020 & Two-stream 2DCNN & V & 6.20 &  7.94 \\ 
    & Deep residual learning \cite{de2019depression} & 2019 & ResNet-50 & V & 6.15 &  8.23 \\ 
    & Multi-channel ensembling \cite{zhou2018visually} & 2018 & Four DCNN & V & 6.21 &  8.39 \\ 
    & Depth-wise video analysis \cite{al2018video} & 2018 & C3D & V & 7.22 & 9.20 \\ 
    & Dual-channel analysis \cite{zhu2017automated} & 2017 & Two DCNN  & V & 7.47 &  9.55 \\ 
    & Feature embedding based network \cite{jan2017artificial} & 2017 & VGG + FDHH &  V & 6.68 & 8.04 \\ 
    & Ensembled feature extraction \cite{kaya2014ensemble} & 2014 & LGBP-TOP + LPQ &  AV & 8.20 & 10.27 \\ 
    & Base paper \cite{valstar2014avec} & 2014 & OpenSMILE + LPQ & AV & 8.86 & 10.86 \\ 
    \bottomrule 
    \end{tabular}
} \vspace{-.15in}
\end{table*}

\begin{table*}[!h]
\centering
\caption{Performance summary of engagement analysis approaches, sorted first by year and then by mean absolute error (`MAE'). Here, `Acc' denotes accuracy and the acronyms used include TCN (Temporal Convolutional Network), LSTM (Long Short-Term Memory), GAP (Gaze-AU-Pose), LBP-TOP (Local Binary Patterns from Three Orthogonal Planes), Deep Facial Spatiotemporal Network (DFSTN), S-WL (Sampling and weighted loss) and LRCN (Long-Term Recurrent Convolutional Network).\vspace{-.07in}}
\label{tab:performance_summary_engagement}
\resizebox{.98\textwidth}{!}
{
    \begin{tabular}{|C{.8cm}|p{4.8cm}|C{.8cm}|p{4.1cm}|C{.9cm}||C{2.5cm}|C{.6cm}|}\toprule
    \textbf{Dataset} & \textbf{Method} & \textbf{Year}& \textbf{Architecture}& \textbf{Modality}& \textbf{MAE} & \textbf{Acc}\\ \midrule 
    \multirow{11}{*}{\rotatebox{90}{\href{https://sites.google.com/view/emotiw2018}{EngageWild} \cite{kaur2018prediction}}} 
    & Facial feature fusion \cite{singh2023have} & 2023 & Transformer based fusion encoder & V & 0.0820 & \textendash\\
    & Sequence embedding optimization \cite{copur2022engagement} & 2022 & Multi-task training  & V & 0.0427 & \textendash\\
    & Real-time CNN classification \cite{savchenko2022classifying} & 2022 & EfficientNet-B0 + Ridge regression & V &  0.0563 & \textendash \\
    & Real-time CNN classification \cite{savchenko2022classifying} & 2022 & EfficientNet-B2 + Ridge regression & V & 0.0702 & \textendash\\
    & Real-time CNN classification \cite{savchenko2022classifying} & 2022 & MobileNet + Ridge regression & V & 0.0722 & \textendash\\
    & Affective ordinal classification \cite{abedi2021affect} & 2021 & Clip-level features + TCN & V &  0.0508 & \textendash\\ 
    & Attention-based hybrid model \cite{zhu2020multi} & 2020 & Attention-based GRU hybrid net & V & 0.0517 & \textendash\\ 
    & Attention-based hybrid model \cite{zhu2020multi} & 2020 & VGG & V & 0.0653 & \textendash\\ 
    & Multi-segment feature analysis \cite{huynh2019engagement} & 2019 & LSTM + FC layers & V &0.0572 & \textendash\\
    & Bootstrap ensemble learning \cite{2019Wang} & 2019 & OpenPose + LSTM & V & 0.0717 & \textendash\\ 
    & Cluster-attention ensemble \cite{chang2018} & 2018 & Attention-based NN & V &0.0441 & \textendash\\ 
    & Sequential behaviour analysis \cite{niu2018} & 2018 & GAP + LBP-TOP & V & 0.0569 & \textendash \\ 
    & Segment-level feature analysis \cite{thomas2018} & 2018 & Dilated-TCN & V  &  0.0655 & \textendash\\ 
    & Feature learning \cite{niu2018} & 2018 & Gaze-AU-Pose - GAP & V & 0.0671 & \textendash\\ 
    & Base paper \cite{kaur2018prediction} & 2018 & OpenFace + LSTM & V & 0.1000 & \textendash\\
    \hdashline
    \multirow{11}{*}{\rotatebox{90}{\href{https://people.iith.ac.in/vineethnb/resources/daisee/index.html}{DAiSEE} \cite{gupta2016daisee}}} 
    & EngageFormer \cite{mandia2025transformer} & 2025 & Multi-view transformer & PV & \textendash & 63.90\%\\
    & Self-supervised masked autoencoder \cite{zhang2024self} & 2024 & ViT-based facial autoencoder & PV & \textendash & 64.74\%\\
    & Multimodal engagement detection \cite{singh2024visiophysioenet} & 2024 & VisioPhysioENet & PV & \textendash & 63.09\%\\
    & Temporal recognition network \cite{selim2022students} & 2022 & EfficientNet B7 + LSTM & V & \textendash & 67.48\%\\
    & Temporal recognition network \cite{selim2022students} & 2022 & EfficientNet B7 + Bi-LSTM & V & \textendash & 66.39\%\\
    & Temporal recognition network \cite{selim2022students} & 2022 & EfficientNet B7 + TCN & V & \textendash & 64.67\%\\
    & Affective ordinal classification \cite{abedi2021affect} & 2021 & Ordinal TCN & V & \textendash & 67.40\%\\
    & Dynamic engagement classifier \cite{abedi2021improving} & 2021 & ResNet + TCN & V & \textendash & 63.90\%\\
    & Dynamic engagement classifier \cite{abedi2021improving} & 2021 & ResNet + LSTM & V & \textendash & 61.15\%\\
    & Dynamic engagement classifier \cite{abedi2021improving} & 2021 & C3D + TCN & V & \textendash & 59.97\%\\
    & Spatiotemporal engagement detector \cite{liao2021deep} & 2021 & DFSTN & V & \textendash & 58.84\%\\
    & Dynamic engagement classifier \cite{abedi2021improving} & 2021 & ResNet +  TCN (S-WL) & V & \textendash & 53.70\%\\ 
    & Class balanced I3D Model \cite{zhang2019annovel} & 2019 & Inflated 3D CNN & V & \textendash & 52.35\%\\ 
    & Base paper \cite{gupta2016daisee} & 2016 & C3D LRCN & V & \textendash & 57.90\%\\
    \bottomrule  
    \end{tabular}
}\vspace{-.05in}
\end{table*}

\subsubsection{Discussion of State-of-the-Art}
Recent advancements in the fields of stress, depression and engagement analysis have increasingly leveraged deep multimodal fusion techniques to accurately capture subtle behavioral and emotional cues. In stress detection using the WESAD dataset, innovative methods using ConvNeXt and GAN based augmentations have achieved high accuracies, while models such as PhysioFormer have demonstrated exceptional performance through advanced feature extraction techniques \cite{li2025multimodal, wang2024physioformer}. For depression analysis on the AVEC benchmarks, techniques that integrate bidirectional fusion with sophisticated spatiotemporal approaches have refined feature detection and significantly reduced error rates \cite{niu2025depression, wang2025automatic}. Recent trends in depression analysis indicate notable improvements in error reduction, underscoring the effectiveness of advanced multimodal fusion strategies in capturing subtle depressive cues. In engagement analysis, the EngageWild dataset and DAiSEE have long served as widely used resources; however, following the introduction of EngageNet in 2023 \cite{singh2023engagenet}, which offers enhanced annotation precision and improved experimental robustness, the community has increasingly shifted toward adopting EngageNet over EngageWild, with transformer based models on DAiSEE further expanding the methodological landscape \cite{mandia2025transformer}.

\section{Applications of Stress, Depression and Engagement Analysis}\label{sec:applications}
\begin{figure*}[]
  \centering
  \includegraphics[width=1\textwidth]{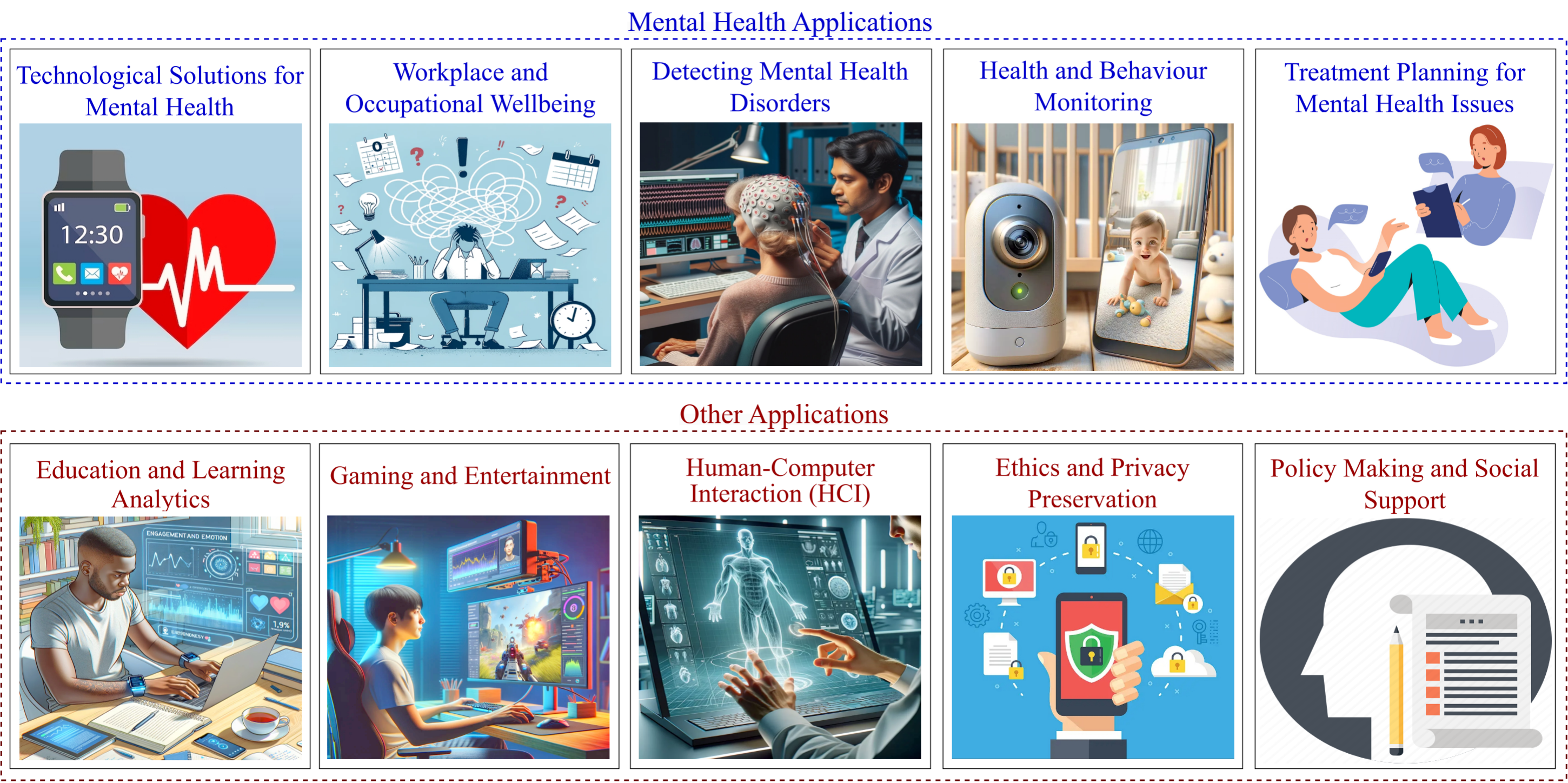}\vspace{-.1in}
  \caption{Mental health and other applications of the computational analysis of stress, depression and engagement. These images were created with \href{https://openai.com/dall-e-2}{DALL·E 2}.\vspace{-.1in}}
  \label{fig:applications}
\end{figure*}

The applications of stress, depression and engagement analysis in mental health and other areas are described in Fig. \ref{fig:applications} and below.\vspace{-.05in} 

\subsection{Mental Health Applications}
\subsubsection{Workplace and Occupational Well-being}\label{sec:WOW-app}
\begin{itemize}[itemsep=1pt]
    \item \textit{Transport and Drivers' Safety}: By analyzing signals such as heart rate and driving behaviour, researchers have developed methods to alert drivers about their stress levels \cite{rastgoo2018critical, nvemcova2020multimodal}. These systems detect stress in real-time, offering suggestions like adjusting cabin settings or applying brakes to enhance safety \cite{siam2023automatic,mou2021driver}, thereby improving vehicle control during stressful situations \cite{hong2018classification}.
    
    \item \textit{Health Professionals' Well-being}: Computational tools are used to understand the ways in which workplace dynamics can influence employee engagement \cite{vedernikov2024tcct}. This analysis aids in the development of enhanced mental health strategies tailored for unusual or atypical situations. For example, researchers have conducted detailed studies on the impact of stress stemming from remote work during the COVID-19 crisis \cite{galanti2021work}.
    
    \item \textit{Social Workers' Mental Health}: Research indicates that high job demands often result in stress and burnout among social workers \cite{travis2016m}. It has also linked these demands to their engagement and mental health, guiding the creation of strategies to help them manage stress \cite{upadyaya2016job}.
    
    \item \textit{Online Meetings:} The significance of computational analysis in online meetings is on the rise within the remote work environment \cite{salam2022automatic}. In this context, the analysis of webcam videos provides instant feedback on attentiveness, aiding in the adaptation of meeting methods \cite{lee2022predicting}. Additionally, stress during online meetings has been analyzed using remote physiological signals and behavioural features \cite{sun2022estimating}.
\end{itemize}

\subsubsection{Detecting Mental Health Disorders}\label{sec:DMHD-app}
\begin{itemize}[itemsep=1pt]
    \item \textit{Anxiety and Stress Detection}: Computational methods are increasingly being applied to detect and evaluate mental health conditions. Video-based facial analysis has been used to assess anxiety symptoms \cite{pampouchidou2020automated}. Additionally, wearable technology has been used to measure physiological responses to stress \cite{anusha2019electrodermal, migovich2024stress}, while ECG and EMG signals have been analyzed for stress detection.
    
    \item \textit{Depression Screening and Suicide Prevention}: Computational analysis has enabled early detection of depressive and suicidal tendencies through monitoring social media, which aids in timely interventions \cite{ragheb2021negatively}. It also supports suicide prevention by identifying risk factors in adolescents engaged with online programs \cite{soares2022effects} and suggests that measuring life satisfaction may predict late-life depression and suicide attempts \cite{o2023life}.
    
    \item \textit{Detection of Post-Traumatic Stress Disorder (PTSD)}: Innovative tools for detecting engagement during vagal nerve stimulation therapy show promise for treating stress from traumatic events \cite{gurel2020automatic,dia2024paying,wang2024application}. Additionally, examining social connections post-stroke reveals their impact on depression and physical impairment, suggesting therapeutic strategies \cite{xezonaki2020affective}.
\end{itemize}

\subsubsection{Health and behaviour Monitoring}\label{sec:HBM-app}
\begin{itemize}[itemsep=1pt]
    \item \textit{Elderly Care}: Stress, depression and engagement analysis techniques facilitate continuous monitoring of the elderly’s physical and emotional states. They have been used to aid in distress and disease prediction \cite{nvemcova2020multimodal, jan2017artificial}. Additionally, wearable technology is used to improve cognitive training in older adults \cite{delmastro2020cognitive}.
    
    \item \textit{Infant Monitoring}: The advancements in Child–Robot Interaction demonstrate that robots, like the `Mio Amico' robot, can adapt to children's engagement levels \cite{filippini2021facilitating}. They can continuously monitor infants and suggest urgent actions. 
    
    \item \textit{Addiction Monitoring}: The computational techniques for stress and engagement analysis facilitate the monitoring of addiction-related behaviours \cite{gupta2023facial}. For instance, analyzing behavioural responses during food consumption offers insights into addiction and emotional eating patterns \cite{casado2023depression}.
\end{itemize}

\subsubsection{Treatment Planning for Mental Health Issues}\label{sec:TPMHD-app}
\begin{itemize}[itemsep=1pt]
    \item \textit{Mental Health Assessment}: Emotion assessment, essential to mental health treatment planning, is supported by facial expression analysis and emotion recognition \cite{gupta2023multimodal}. It utilizes clinical interviews, behavioural analysis and social media analysis approaches \cite{mallol2019hierarchical}.
    
    \item \textit{Therapeutic Interventions and Counseling}: Understanding stress, depression and engagement provides valuable real-time feedback, thereby improving the effectiveness of therapeutic interventions and counselling sessions \cite{banerjee2023heart, mou2021driver}.
    
    \item \textit{Personalized Coping Mechanisms}: To aid individuals experiencing challenging emotional states, personalized coping strategy systems have been developed \cite{zhu2023stress, li2023mha}. These systems offer tailored support to users facing various forms of emotional distress.
    
    \item \textit{Treating Cognitive Degeneration Disorders}: Computational methods and wearable sensors help older adults with cognitive training and stress detection \cite{delmastro2020cognitive, carrizosa2021systematic}. These methods use small sensors to track the heart rate and body movement. By analyzing the data from these sensors, researchers can monitor the cognitive load and stress levels to facilitate training for older adults. 
\end{itemize}

\subsubsection{Implementation Solutions for Mental Health}\label{sec:TSMH-app}
\begin{itemize}[itemsep=1pt]
    \item \textit{Mobile Application Development}: Computational analysis has significantly contributed to personalized mental healthcare via mobile applications. These applications utilize user interaction data, such as typing speed and phone usage, to non-invasively assess stress levels, enabling real-time monitoring and intervention for stress-related conditions \cite{ciman2016individuals, lipschitz2019adoption}.
    \item \textit{Wearable Technology-based Well-being Analysis}: Wearable devices have revolutionized real-time well-being analysis \cite{schmidt2018introducing}. Privacy-preserving stress monitoring is now possible with smartwatches \cite{xu2021privacy}, while wrist devices for electrodermal activity can be used for non-invasive stress detection \cite{anusha2019electrodermal}.
\end{itemize}

\subsection{Other Applications}
\subsubsection{Education and Learning Analytics}\label{sec:ELA-app}
\begin{itemize}[itemsep=1pt]
    \item \textit{Improving Student Engagement}: Computational analysis has improved online learning by detecting student engagement using facial emotion recognition \cite{shen2022assessing, gupta2023facial}. Techniques have been developed to predict mind-wandering during online lectures and to detect students' engagement in classroom environments \cite{lee2022predicting, ts2020automatic}.
    
    \item \textit{Personalized Learning}: Adaptive learning technologies have been developed to improve personalized experiences \cite{lam2019context}. They also enable customized support by utilizing automated processes to meet the unique needs of each learner \cite{cho2017automated}.
\end{itemize}
 
\subsubsection{Gaming and Entertainment}\label{sec:GA-app} 
\begin{itemize}[itemsep=1pt]
    \item \textit{Creating More Engaging Gaming Experiences}: The player engagement in game-based learning can be explored by measuring physiological signals \cite{ober2021detecting}. Games have also been used as therapeutic interventions for depression \cite{fang2023multimodal, niu2020multimodal}.
    
    \item \textit{Improving the Virtual Reality (VR) Experience}: To enhance work engagement and alleviate stress levels, researchers have developed and implemented virtual reality-based games \cite{wiezer2013serious}. These immersive experiences leverage advanced technology to create interactive environments, offering employees an engaging and stress-relieving alternative to traditional methods.
\end{itemize}

\subsubsection{Human-Computer Interaction (HCI)}\label{sec:HCI-app} 
\begin{itemize}[itemsep=1pt]
    \item \textit{Engagement Detection in HCI}: Advanced DL techniques have been employed to measure users' engagement in HCI scenarios using video-based facial expressions and consumer interaction patterns on social media platforms \cite{abedi2021affect}.
    \item \textit{Understanding Human Emotions in HCI}: Advancements in HCI have led to applications such as creating emotion databases, using neural networks for emotion detection and exploring settings like classrooms and gaming to monitor and respond to users’ emotional states \cite{kang2023k}.
    \item \textit{Digital Engagement and Social Media Analysis}: By analyzing online behaviour patterns, researchers can identify social trends \cite{kastrati2023soaring}. Digital engagement analysis can be used to understand public sentiments and mental health aspects \cite{de2021sadness}. 
\end{itemize} 

\subsubsection{Ethics and Privacy Preservation}\label{sec:PE-app} 
Given the heightened risk of data breaches, computational tools are being developed to ethically handle mental health data, enhancing well-being while protecting privacy and rights \cite{de2013predicting}. 

\subsubsection{Policy Making and Social Support}\label{sec:PMSS-app} 
The insights from stress, depression and engagement analysis are useful in developing strategies to address mental health concerns \cite{ashwin2020affective}. These insights are instrumental for governments and policymakers in devising effective social support initiatives \cite{soares2022effects}. 

\subsection{Discussion of Applications}
In applying computational analysis to mental health and related fields, the context dependency of engagement, stress and depression is crucial. Identifying whether an individual is engaged with specific content or a particular person is often more important than assessing engagement alone. For instance, in older adults with dementia, engagement with either a recommender system or a human partner can promote cognitive activation \cite{steinert2022evaluation}, while in human-robot collaborative learning environments, distinguishing whether a learner’s attention is directed toward the robot or a human instructor is essential for effective adaptivity \cite{papadopoulos2016relative}. {Since heightened stress and depression reduce motivation and focus, interventions increasingly target these states, particularly in workplaces and education, to sustain engagement \cite{krishnan2008molecular, maydych2019interplay}.} These examples highlight the need for nuanced, context-aware approaches that optimize user experience and therapeutic efficacy.

\section{Challenges and Future Directions}\label{sec:futuredir}
Computational analysis of stress, depression and engagement uncovers the following challenges and future research avenues.

\begin{itemize}[itemsep=.75pt]
\item \textbf{Emerging Generative AI Approaches}: 
{Advancements in generative AI, such as leveraging large language models (LLMs) or generating synthetic data, have shown promise in various domains. Yet, their application in analyzing stress, depression and engagement is less explored. Recent efforts include generating synthetic health sensor data for stress detection \cite{lange2024generating}, enhancing educational engagement through generative tools \cite{magana2024ai} and assessing LLM capabilities for depression detection \cite{ohse2024zero}. Future research should employ generative AI techniques to further mental health analysis.}

\item \textbf{Lack of Large-scale Datasets}:
The analysis of stress, depression and engagement is hindered by the scarcity of large-scale datasets. Collaborative efforts across disciplines can address this challenge by pooling resources, sharing data and conducting joint analyses within privacy guidelines. Such collaboration enhances dataset quality and availability \cite{lee2022predicting}.

\item \textbf{Data Labeling and Multimodal Data Integration}: 
With the rise of wearable devices and IoT, there are numerous new data sources available. While these offer rich insights, labelling them effectively remains a challenge \cite{mehta2022three}. The research efforts are required to create robust labelling methodologies and integrate the diverse data streams to form a coherent understanding of mental states \cite{tao2023towards}.

\item \textbf{Model Generalization}:
The emergence of stress, depression and engagement varies greatly among individuals, presenting challenges for generalizing computational models \cite{benini2019influence}. Techniques like domain adaptation and multi-task learning offer potential solutions by transferring knowledge between datasets to account for variability in emotional analysis \cite{huang2020domain}. Future efforts should focus on refining these models for improved prediction and response to emotional shifts \cite{psaltis2017multimodal}.

\item \textbf{Dynamic Analysis}:
Emotions are dynamic and subject to change over time. Some emotions can shift very quickly, while others may change more slowly \cite{niedenthal2007embodying}. This variability presents a challenge for computational models, as they must be capable of adapting to and accurately capturing these different rates of emotional changes. It is important to develop computational models that are sensitive to these varying speeds. While current computational models can analyze data sequences to comprehend evolving emotions, additional research is required to enhance their efficacy in monitoring and predicting emotional variations over time \cite{ashwin2020impact}.

\item \textbf{Interpretability of Emotion Understanding Models}: 
The complexity of ML and DL computational models presents challenges in interpreting their internal workings, particularly when analyzing sensitive mental health data \cite{naegelin2023interpretable}. Future research is required to improve the interpretability of these models to ensure their trustworthiness and effective utilization in mental health contexts \cite{zhou2018visually}.

\item \textbf{Individual \& Cultural Differences}: 
The variations in styles used by individuals and cultures to express emotions present challenges in their analysis \cite{ober2021detecting}. For example, diverse skin tones complicate facial analysis for emotion understanding \cite{prasetio2018facial}, while variations in voice annotations affect audio-based emotion analysis \cite{sardari2022audio}. Addressing this challenge requires acquiring data from diverse populations and developing computational systems that can adapt to individual changes over time \cite{ashwin2020impact}.

\item \textbf{Context Dependency}: 
The emotions are not static and can vary widely over time and across different situations. Understanding stress, depression and engagement requires considering the context in which they occur. Constructing models capable of recognizing context is important and it is also essential to design them so that they can adapt to changing contexts in real-time \cite{liao2021deep}.

\item \textbf{Integration of Computational, Psychological, Medical and Social Analysis}: 
Developing computational methods that merge insights from medicine, psychology and sociology is crucial in mental health research \cite{coppersmith2015clpsych}. Standardized approaches are vital for universal application and recognition in these fields, fostering a unified understanding of mental health and facilitating research translation into practice \cite{gurel2020automatic}. Synergizing insights from these disciplines is key to improving the accuracy and effectiveness of mental health interventions.
\end{itemize}

\subsection{Discussion of Challenges and Future Directions} 
{Future research must not only address critical challenges in accuracy and real-world applicability but also capitalize on the dynamic interplay among stress, depression and engagement to develop effective interventions. Mitigating daily stressors can enhance engagement and reduce depressive symptoms, creating a reinforcing cycle that bolsters cognitive and emotional well-being \cite{pizzagalli2014depression}. Tailoring interventions to individual engagement profiles may help sustain motivation under heightened stress. Meanwhile, advancements in generative AI, such as LLMs and synthetic data augmentation, show promise but remain underexplored here \cite{lange2024generating}. Data scarcity demands collaborative efforts to improve dataset quality \cite{lee2022predicting}. Robust labeling and fusion techniques are needed to handle varied modalities \cite{mehta2022three} and model generalization requires domain adaptation for high inter-individual variability \cite{benini2019influence}. Additionally, these states evolve over time, necessitating dynamic, context-aware models that adapt to changing contexts \cite{niedenthal2007embodying}. The black-box nature of ML and DL underscores the need for interpretability and accounting for cultural and contextual factors remains crucial for unified, adaptive mental health analysis \cite{naegelin2023interpretable}.} 

\section{Conclusions}\label{sec:conc}
This survey is the first to collectively review computational methods for detecting stress, depression and engagement. It traces the evolution from traditional techniques to advanced ML and DL approaches, highlighting their potential to improve mental healthcare with early detection, personalized interventions and ongoing monitoring. We explore the complexities of multimodal datasets and the challenges they introduce, highlighting a shift towards sophisticated algorithms that offer deeper mental health insights. Our review underscores DL's transition, emphasizing its accuracy despite its computational intensity and substantial data requirements. The incorporation of multimodal data, including wearables and social media analysis, mirrors the innovative direction of current research and the movement towards interpretable methods for application transparency. This paper highlights the transformative role of computational methods in mental healthcare and calls for ongoing innovation to advance personalized and effective interventions.

\section*{Acknowledgment}
The authors express gratitude to the Center for Machine Vision and Signal Analysis, University of Oulu, Finland for the academic and literary resources provided under the Research Council of Finland Profi 5 HiDyn grant 24630111132. {The work was partially supported by the Eudaimonia Institute of the University of Oulu.}

\ifCLASSOPTIONcaptionsoff
  \newpage
\fi

\bibliographystyle{ieeetr}  
\bibliography{ref} 

\begin{IEEEbiography}[{\includegraphics[width=1in,height=1.25in,clip,keepaspectratio]{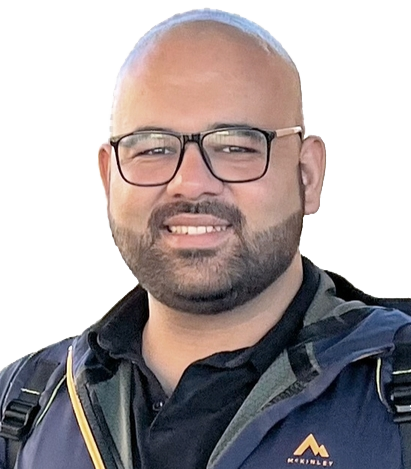}}]{Puneet~Kumar} (Member, IEEE) received his B.E. and M.E. degrees in Computer Science in 2014 and 2018, respectively and his Ph.D. from the IIT Roorkee, India in 2022. He has worked at Oracle, Samsung R\&D and PaiByTwo Pvt. Ltd. and is now a Postdoctoral Researcher at the University of Oulu, Finland. His research interests include Affective Computing, Multimodal and Interpretable AI, Mental Health and Cognitive Neuroscience. He has published in top journals and conferences and received institute medal in M.E., the best thesis award and several best paper awards. For more information, visit his webpage at \url{www.puneetkumar.com}.\vspace{-.47in}
\end{IEEEbiography}

\begin{IEEEbiography}[{\includegraphics[width=1in,height=1.25in,clip,keepaspectratio]{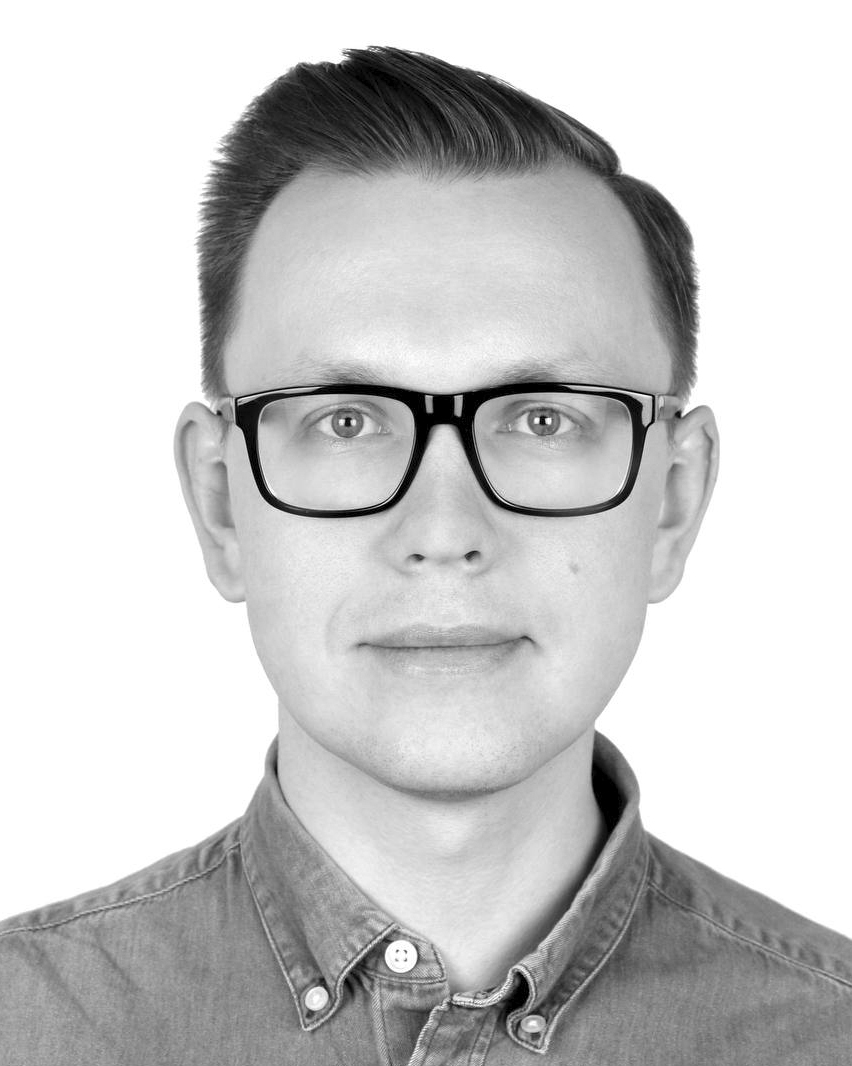}}]{Alexander~Vedernikov} (Member, IEEE) received his M.Sc. from Politecnico di Milano, Italy, in 2018 and his Ph.D. in Mathematics from Skolkovo Institute of Science \& Technology, Moscow, Russia, in 2022. He is currently with the Center for Machine Vision and Signal Analysis at the University of Oulu, Finland, working in Affective Computing, Facial Analysis and Emotion Understanding. For more details, visit his webpage at \href{https://www.linkedin.com/in/a-vedernikov}{www.linkedin.com/in/a-vedernikov}.\vspace{-.57in}
\end{IEEEbiography}

\begin{IEEEbiography}[{\includegraphics[width=1in,height=1.25in,clip,keepaspectratio]{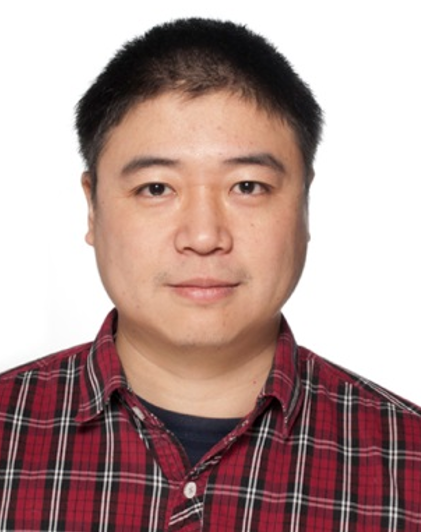}}]{Yuwei~Chen} received B.E. and M.Sc. from Zhejiang University, China (1999, 2002) and Ph.D. in Circuits and Systems from the Chinese Academy of Sciences (2005). He later obtained a Doctor of Tech. in Telecommunication Software from Aalto University, Finland (2020). He contributed to China's first moon-exploration satellite, Chang’e, developing its echo-detecting laser range sensor and prototyped China's first airborne pushbroom laser scanner. Currently, he is director general at Advanced Laser Technology Anhui and guest professor at the Chinese Academy of Sciences and Zhejiang University. He works on hyperspectral LiDAR, radar and navigation and has published over 200 papers and 16 patents. For more information, visit his profile at \url{www.researchgate.net/profile/Yuwei-Chen}.\vspace{-.47in}
\end{IEEEbiography}

\begin{IEEEbiography}[{\includegraphics[width=1in,height=1.25in,clip,keepaspectratio]{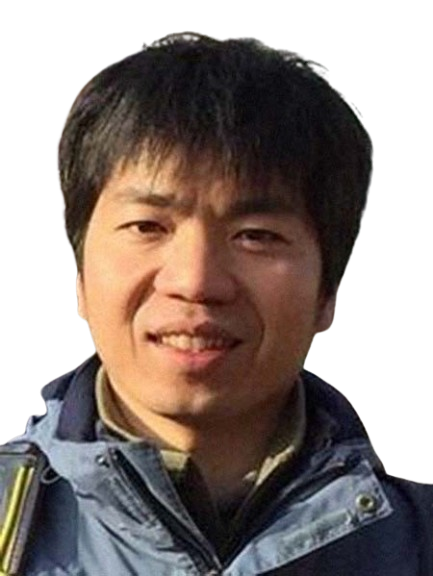}}]{Wenming~Zheng} (Senior Member, IEEE) received B.S. in Computer Science from Fuzhou University (1997), M.S. from Huaqiao University (2001) and Ph.D. in Signal Processing from Southeast University (2004). Since then, he is with the Research Center for Learning Science, Southeast University. He is a professor at the School of Biological Science and Medical Engineering and the Key Laboratory of Child Development and Learning Science, Ministry of Education. He works on affective computing, pattern recognition, machine learning and computer vision. He is an Associate Editor for IEEE Transactions on Affective Computing and IEEE Transactions on Cognitive and Developmental Systems. For more information, visit his profile at \url{https://ieeexplore.ieee.org/author/37273477300}.\vspace{-.47in}
\end{IEEEbiography}

\begin{IEEEbiography}[{\includegraphics[width=1in,height=1.25in,clip,keepaspectratio]{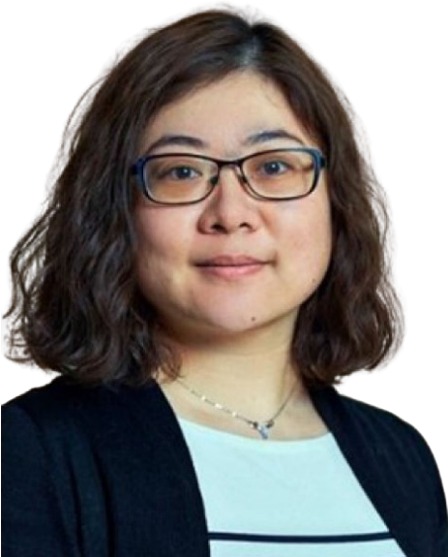}}]{Xiaobai~Li} (Senior Member, IEEE) received her B.Sc. and M.Sc. degrees in 2004 and 2007, respectively and her Ph.D. from the University of Oulu, Finland, in 2017. She is a faculty member at the State Key Lab of Blockchain and Data Security, Zhejiang University, China and an Adjunct Professor at the University of Oulu. Her research focuses on Affective Computing, Facial Expression Recognition, Micro-Expression Analysis and Remote Physiological Signal Analysis. She has co-chaired international workshops at CVPR, ICCV, FG and ACM MM and serves as an Associate Editor for IEEE Transactions on Circuits and Systems for Video Technology, Frontiers in Psychology and Image and Vision Computing. For more information, visit her webpage at \url{https://xiaobaili-uhai.github.io/}. 
\end{IEEEbiography}

\end{document}